\documentclass[aps,twocolumn,amsmath,amssymb,superscriptaddress]{revtex4-1}

\usepackage{amssymb,amsfonts,amsmath}
\usepackage{graphicx}
\usepackage{epsfig}
\usepackage{xcolor}
\usepackage{ulem}
\usepackage{SIunits}
\usepackage{epstopdf}
\usepackage[utf8]{inputenc}
\usepackage[english]{babel}
\usepackage{ulem}
\graphicspath{ {./Bilder/} }

\definecolor{darkblue}{rgb}{0, 0, 0.8}
\usepackage[colorlinks=true, breaklinks=true, linkcolor=darkblue, citecolor=darkblue, urlcolor=darkblue]{hyperref}

\newcommand{\rs}[1]{\rm{\scriptscriptstyle #1}}

\begin{document}

\title{Experimental realization of a symmetry protected topological phase \\ of interacting bosons with Rydberg atoms
}
 	\author{Sylvain de L\'es\'eleuc}
	\email[The four authors contributed equally to this work.]{}
 	\affiliation{Laboratoire Charles Fabry, Institut d'Optique Graduate School, CNRS, Universit\'e Paris-Saclay, 91127 Palaiseau Cedex, France}
	\author{Vincent Lienhard}
	\email[The four authors contributed equally to this work.]{}
 	\affiliation{Laboratoire Charles Fabry, Institut d'Optique Graduate School, CNRS, Universit\'e Paris-Saclay, 91127 Palaiseau Cedex, France}
	\author{Pascal Scholl}
	\affiliation{Laboratoire Charles Fabry, Institut d'Optique Graduate School, CNRS, Universit\'e Paris-Saclay, 91127 Palaiseau Cedex, France}
	 \author{ Daniel Barredo}
	\affiliation{Laboratoire Charles Fabry, Institut d'Optique Graduate School, CNRS, Universit\'e Paris-Saclay, 91127 Palaiseau Cedex, France}
	\author{Sebastian~Weber }
	\email[The four authors contributed equally to this work.]{}
	\affiliation{Institute for Theoretical Physics III and Center for Integrated Quantum Science and Technology, University of Stuttgart, 70550 Stuttgart, Germany}
	\author{Nicolai Lang }
	\email[The four authors contributed equally to this work.]{}
	\affiliation{Institute for Theoretical Physics III and Center for Integrated Quantum Science and Technology, University of Stuttgart, 70550 Stuttgart, Germany}
	\author{Hans Peter B\"uchler}
	\affiliation{Institute for Theoretical Physics III and Center for Integrated Quantum Science and Technology, University of Stuttgart, 70550 Stuttgart, Germany}
	\author{Thierry Lahaye}
	\affiliation{Laboratoire Charles Fabry, Institut d'Optique Graduate School, CNRS, Universit\'e Paris-Saclay, 91127 Palaiseau Cedex, France}
	\author{Antoine Browaeys}
	\affiliation{Laboratoire Charles Fabry, Institut d'Optique Graduate School, CNRS, Universit\'e Paris-Saclay, 91127 Palaiseau Cedex, France}
\date{\today}

\begin{abstract}

The concept of topological phases is a powerful framework to characterize ground states of quantum many-body systems that goes beyond the paradigm of symmetry breaking. While a few topological phases appear in condensed matter systems, a current challenge is the implementation and study of such quantum many-body ground states in artificial matter. Here,  we report the experimental realization of a symmetry protected topological phase of interacting bosons in a one-dimensional lattice, and demonstrate a robust ground state degeneracy attributed to protected edge states. The setup is based on atoms trapped in an array of optical tweezers and excited into Rydberg levels, which gives rise to hard-core bosons with an effective hopping by dipolar exchange interaction.  
\end{abstract}

\maketitle

\section*{Introduction}

The paradigm of symmetry breaking has proven very successful for characterizing quantum phases. However, not all of them follow this paradigm, and some of these phases are nowadays characterized in the framework of topological phases. 
The most prominent example of a topological phase is the integer quantum Hall state with its remarkably robust edge states giving rise to a quantized Hall conductance~\cite{Klitzing1980}. For a long time, it was believed that such phases occurred only in the presence of a magnetic field, until the prediction of topological insulators~\cite{Kane2005a} revealed a novel class of topological states of matter, nowadays denoted as symmetry protected topological phases (SPT). They occur in systems displaying an excitation gap in the bulk, i.e., bulk insulators, and an invariance under a global symmetry. Their defining property is that the ground state at zero-temperature cannot be transformed into a conventional  insulating state upon deformations of the system that do not close the excitation gap or violate the symmetry. In particular, the edge states are robust to any perturbation commuting with the symmetry operators. 

SPT phases were first predicted and observed in materials where the interaction between electrons can be effectively neglected~\cite{Hasan2010,Xiao2011}. In this specific case of non-interacting fermions, SPT phases can be classified based on the action of the Hamiltonian on a single particle~\cite{Chiu2016,Ludwig2016}. Thus, the appearance of robust edge states is  fully understood from the single-particle eigenstates. This remarkable simplification motivated experimental studies of topological phenomena at the single-particle level with artificial quantum matter realized on ultracold atoms platforms~\cite{Jotzu_2014,Flaschner_2016,Aidelsburger_2014,Stuhl_2015,Mancini_2015,Lohse2015,Tai2017}, and in classical systems of coupled mechanical oscillators~\cite{Susstrunk2015,Nash2015}, as well as optical~\cite{Hafezi2013,Rechtsman2013,Ling2015,Lu2016} or radio-frequency circuits~\cite{Ningyuan2015}, and plasmonic systems~\cite{St-Jean2017,Bandres2018}.


\begin{figure*}[htb]
	\includegraphics[width=1.3\columnwidth]{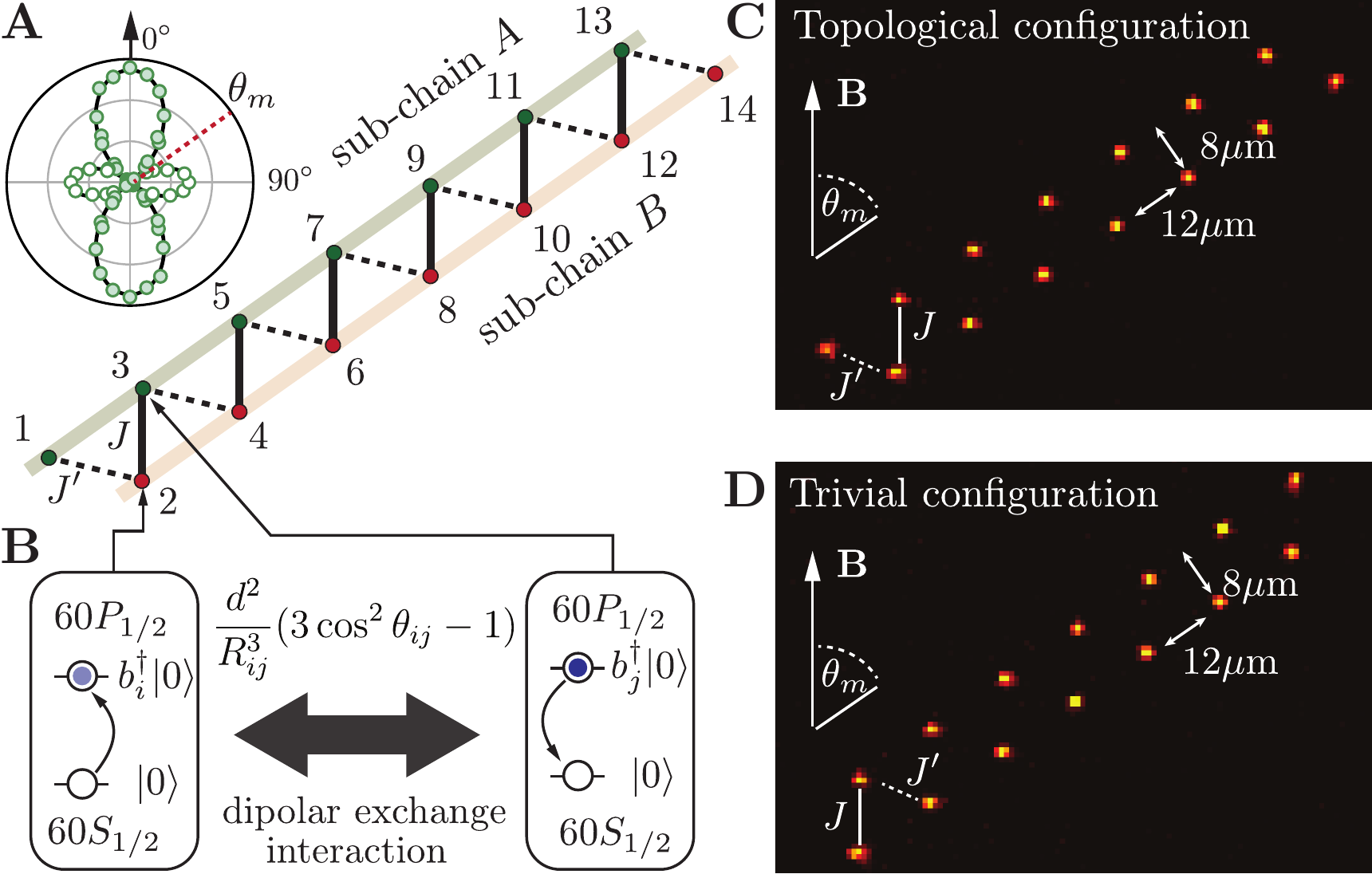}
	\caption{ {\bf Bosonic SSH model}. (A) Dimerized one-dimensional lattice and the two sub-lattices $A$ and $B$. The staggered nearest neighbor hopping energies are denoted as $J$ and $J'$ with $|J|> |J'|$. (B) Each lattice site hosts a Rydberg atom with two relevant levels: $60 S_{1/2}$ being the vacuum state $|0\rangle$ and $ 60 P_{1/2}$ describing a bosonic particle $b_{i}^{\dag}|0\rangle$. The dipolar exchange interaction provides a hopping of the particles.  (Inset in A) Angular dependence of the hopping amplitude measured between two sites; filled (empty) disk: positive (negative) amplitude. It vanishes and changes sign at the angle $\theta_m \simeq 54.7^\circ$. The solid line is the theoretical prediction. Error bars, denoting the standard error of the mean (s.e.m.), are smaller than the symbol size. (C-D) Single-shot fluorescence images of the atoms assembled in the artificial structure for the topological  (C) and the trivial (D) configuration. The chain is tilted by the angle $\theta_m$ to cancel couplings between sites in the same sub-lattice. 
		\label{fig2}}
\end{figure*}

In contrast, the situation is different for bosonic SPT phases as the ground state of non-interacting bosons is a Bose-Einstein condensate. Therefore, it is well established that strong interactions between the particles are required for the appearance of topological phases. Their classification is not derived from single-particle properties, but requires the analysis of the quantum many-body ground state; a classification of bosonic interacting SPT phases based in terms of group cohomology has been achieved~\cite{Chen2012}. A notable example is the Haldane phase of the anti-ferromagnetic spin-1 chain~\cite{Haldane1983}, which has been experimentally observed in some solid-state materials \cite{Hagiwara90,Glarum91}. However, the realization of topological phases in artificial matter, where one has full microscopic control on the particles, would allow to gain a deeper understanding on the nature of such topological states of matter. A first step has recently been achieved by introducing interactions between bosonic particles in a system with a topological band structure~\cite{Tai2017,Liberto2016}, but studies were restricted to the two-body limit, still far from the many-body regime.

Here, we report the first realization of a many-body SPT phase of interacting bosonic particles in an artificial system. Our setup is based on a staggered one-dimensional chain of Rydberg atoms, each restricted to a two-level system, resonantly coupled together by the dipolar interaction~\cite{Browaeys2016}. We use this to encode hard-core bosons, i.e., bosonic particles with infinite on-site interaction energy, coherently hopping along the chain. The system then realizes a bosonic version of the Su-Schrieffer-Heeger (SSH) model~\cite{SSH1979}; the latter originally described fermionic particles hopping on a dimerized lattice, giving rise to a SPT phase of non-interacting fermions. Similarly, our bosonic setup gives rise to two distinct phases of an half-filled chain: a trivial one with a single ground state and an excitation gap, and a SPT phase, with a four-fold ground state degeneracy due to edge states, and a bulk excitation gap. Following an adiabatic preparation of a half-filled chain, we detect the ground state degeneracy in the topological phase and probe the zero-energy edge states. Furthermore, we experimentally demonstrate the robustness of the SPT phase under a perturbation respecting the protecting symmetry, and show that this robustness cannot be explained at the single-particle level, a feature that distinguishes our system from non-interacting SPT phases.

\section*{SSH model for hard-core bosons}

The SSH model is formulated on a one-dimensional lattice with an even number of sites $N$ and staggered hopping of particles, see Fig.~\ref{fig2}A. It is convenient to divide the lattice into two sub-lattices: $A=\{1,3, \ldots, N-1\}$, involving odd lattice sites, and $B= \{2,4, \dots, N\}$,  with even sites. Then, a particle on site $i$ of one sub-lattice can hop to a site $j$ of the other sub-lattice with a hopping amplitude $J_{ij}$ (we do not restrict the system to nearest neighbor hopping). The many-body Hamiltonian is 
\begin{equation}
 H = -\!\!  \sum_{i \in A, j \in B} J_{i j} \left[ b^{\dag}_{i} b_{j}^{}+b^{\dag}_{j} b_{i}^{} \right],
\label{eq:hamiltonian}
\end{equation}
with $b^\dagger_i$ ($b_i^{}$) the creation (annihilation) operator of a particle on site $i$. In the original formulation of the SSH model, the particles are non-interacting fermions. Here we consider hard-core bosons and the operators $b^\dagger_i$ ($b_i^{}$)  satisfy bosonic commutation relations on different sites $i \neq j $, and additionally the hard-core constraint $(b_i^\dagger)^2 = 0$, as two bosons cannot occupy the same site~$i$. In our realization, the nearest neighbor hoppings are dominant with their energies denoted as $J_{2i,2i+1} = J$ and $J_{2i-1,2i} = J'$ with $|J'| < |J|$, and are sufficient to describe the qualitative behavior of the model.

\begin{figure*}[t!]
  \includegraphics[width=\textwidth]{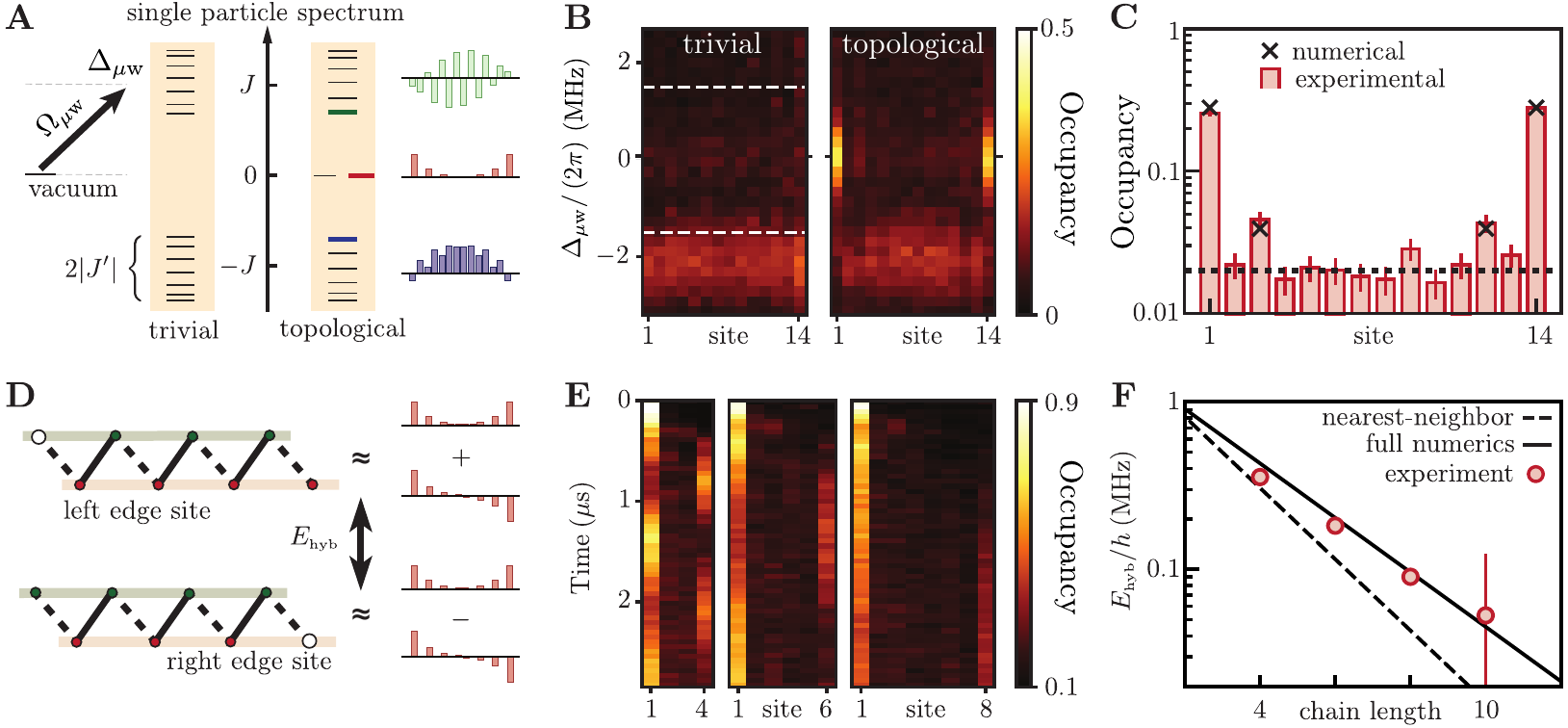}
  \caption{{\bf Single particle properties}. (A) Single-particle spectrum for the trivial and the topological configuration probed by microwave spectroscopy. Right: selection of single-particle wavefunctions. (B) Experimental site-resolved spectra showing the averaged occupancy of each site as a function of $\Delta_{\mu \mathrm{w}}$. The lower band bulk states are always observed, whereas the upper band is not visible due to a suppressed coupling to the microwave probe. Edge states at zero energy appear only for the topological configuration. The white dashed lines indicate the calculated gap. (C) Spatial distribution of the edge states, observed (red bars) and calculated (black crosses), showing an exponential localization on the edges. The dashed line indicates the 2~\% noise level caused by preparation and detection errors. (D) Particle transfer between the two edges: a particle on the left edge is essentially a superposition of the symmetric and anti-symmetric zero-energy modes, split in energy by $E_{\rm hyb}$ due to the hybridization for finite chains. (E) Observation of the transfer for chains of $N=$ 4, 6 and 8 sites after injecting a particle on the leftmost site. (F) From the transfer frequency, we obtain the hybridization energy $E_{\rm hyb}$ (red disks) and compare it to calculations keeping only nearest neighbor hoppings (dashed line) and including the full dipolar interaction (solid line). 
  \label{fig3}}
\end{figure*}

At the single particle level, the spectrum of the Hamiltonian in Eq.~(\ref{eq:hamiltonian}), shown in Fig.~\ref{fig3}A, is obtained by diagonalizing the coupling matrix $J_{ij}$. It displays two bands separated by a spectral gap $2(|J|-|J'|)$ and, depending on the boundaries of the chain, localized zero-energy edge modes. There are two such modes for a chain ending with weak links $J'$ (topological configuration, Fig.~\ref{fig2}C) and none if the chain ends with strong links $J$ (trivial configuration, Fig.~\ref{fig2}D). The topology of the bands emerges from the sub-lattice (or chiral) symmetry of the SSH Hamiltonian~\cite{Chiu2016,Ludwig2016}, which notably constrains the hopping matrix $J_{i j}$ to connect only sites of different sub-lattices, e.g., next nearest neighbor hoppings $J_{i, i+2} = J''$ are forbidden. The existence and degeneracy of edge modes are topologically protected from any perturbation that does not break the sub-lattice symmetry. These single-particle properties of the coupling matrix $J_{ij}$ defining the SSH model have been observed in many platforms such as, e.g., ultracold atoms~\cite{Atala2013, Meier2016}, polaritons in array of micropillars~\cite{St-Jean2017} or mechanical granular chains~\cite{Chaunsali2017}.

We now turn to the properties of the quantum many-body ground state corresponding to an half-filled chain. For non-interacting fermions, the properties of the SSH chain follows from the Fermi sea picture based on the single-particle eigenstates:  in the trivial configuration one obtains a single insulating ground state, while, in the topological configuration, (i) an excitation gap appears in the bulk, and (ii) the ground state is four-fold degenerate as the two zero-energy edge modes can be either empty or occupied. For interacting bosons, the description of the many-body ground state is much more challenging. In the special case with only nearest neighbor hoppings $J$ and $J'$, the bosonic many-body ground states for hard-core bosons can be derived via a Jordan-Wigner transformation from the fermionic ones, and inherits the properties of a bulk excitations gap and a four-fold ground state degeneracy in the topological configuration~\cite{suppl_mat}. Based on the general classification of bosonic SPT phases~\cite{Chen2012}, these properties are robust by adding longer-range hoppings as well as additional interactions between the bosons as long as the bulk gap remains finite and the protecting symmetry is respected. Here, the protecting symmetries are the particle number conservation and the anti-unitary operator
\begin{equation}
\mathcal{S}_{\rs B} = \prod_{i=1}^{N} \left[b_{i}^{}+ b_{i}^{\dag}\right] K,
\label{symmetry}
\end{equation}
where $b_{i}^{}+ b_{i}^{\dag}$ is a particle-hole transformation and $K$ denotes the complex conjugation. In contrast to the chiral symmetry (protecting the single-particle properties and the fermionic SPT phase), next nearest neighbor hoppings  $J''=J_{i, i+2}$ are symmetry allowed, i.e., $\left[H,\mathcal{S}_{\rs B} \right] = 0$ even when including $J''$. We will demonstrate this fundamental difference by engineering a perturbation which shifts the edge modes away from zero-energy at the single particle level, but preserves the ground state degeneracy in the bosonic many-body system.

\section*{Experimental realization with Rydberg atoms}

Our realization of the bosonic SSH model is performed on an artificial structure with $N = 14$ sites of individually trapped $^{87}$Rb atoms~\cite{Kim2016,Endres2016,Barredo2016} (Fig.~\ref{fig2}C-D). The motion of the atoms is frozen during our experiment, occurring in a few microseconds. Coupling between the atoms is achieved, despite the large inter-atomic distance ($\sim 10 \, \mu$m), by considering Rydberg states for which the dipole-dipole coupling is enhanced to a few MHz~\cite{Saffman2010,Browaeys2016}. 

We first prepare each atom in a Rydberg s-level,  $|60 S_{1/2}, m_J = 1/2 \rangle$, using a two-photon stimulated Raman adiabatic passage (STIRAP) with an efficiency of 95\%~\cite{suppl_mat}. From there, the atom can be coherently transferred to a Rydberg p-level,  $|60 P_{1/2}, m_J = -1/2 \rangle$, using a microwave field tuned to the transition between the two Rydberg levels ($E_0/h \sim 16.7$ GHz). The detuning from the transition is $\Delta_{\mu \mathrm{w}}$, and the Rabi frequency is $\Omega_{\mu \mathrm{w}}/(2 \pi) \sim 0.1 - 20$~MHz. We denote the state with all Rydberg atoms in the s-level as the `vacuum' $|0\rangle$ of the many-body system, while a Rydberg atom at site $i$ excited in a p-level is described as a bosonic particle $b^{\dag}_{i}|0\rangle$. Since each Rydberg atom can only be excited once to the p-level, we obtain naturally the hard-core constraint. The resonant dipolar interaction occuring between the s- and p-levels of two Rydberg atoms at site $i$ and $j$ gives rise to hopping of these particles~\cite{Browaeys2016}, as illustrated in Fig.~\ref{fig2}B. We use this to engineer the hopping matrix $J_{ij}$. At the end of the experiment, we de-excite atoms in the Rydberg s-level to the electronic ground state and detect them by fluorescence imaging, while an atom in the Rydberg p-level is lost from the structure. The detection errors are at the few percent level~\cite{suppl_mat}. For each experimental run, we thus obtain the occupancy of each site, which is then averaged by repeating the experiment every $\sim 0.3 $~s.

In order to implement the sub-lattice symmetry, we use the angular dependence of the dipolar coupling $J_{ij}=d^2 (3\cos^2\theta_{i j}\!-\!1)/R_{i j}^3$ with $d$ the transition dipole moment between the two Rydberg levels.
The hopping depends on the separation $R_{i j}$, as well as the angle $\theta_{i j}$ with respect to the quantization axis defined by the  magnetic field $B_{z} \simeq 50$~G. In Fig.~\ref{fig2}A, we show the measured angular dependence, vanishing at the `magic angle' $\theta_{m} = \arccos(1/\sqrt{3})\approx 54.7^{\circ}$, which allows us to suppress the hopping along this direction. By arranging the atoms in two sub-chains aligned along the magic angle, we thus satisfy the sub-lattice symmetry. The measured nearest neighbor couplings are $J/h = 2.42(2)$~MHz and $J'/h = -0.92(2)$~MHz, in full agreement with numerical determination of the pair potential~\cite{Weber2017}. The dipolar interaction also gives rise to longer range hopping on the order of $\sim  0.2$~MHz to third neighbors, which do not qualitatively change the properties of our system but are fully taken into account for quantitative comparison between theory and experiment.

\section*{Single-particle spectrum}

As a benchmark of our system, we first study the properties of a single particle in the chain. The single-particle spectrum is probed by microwave spectroscopy (Fig.~\ref{fig3}A). Initializing the vacuum state $|0\rangle$ with all Rydberg atoms in the s-level, a weak microwave probe with a Rabi frequency $\Omega_{\mu \mathrm{w}}/(2 \pi) = 0.2$~MHz applied for a time $t = 0.75 \, \mu$s can lead to the coherent creation of a particle only if an eigenstate energy matches the microwave detuning $\Delta_{\mu \mathrm{w}}$ and if this state is coupled to $|0\rangle$ by the microwave field. We show in Fig.~\ref{fig3}B the site-resolved probability to find a particle on a given site for the two different chain configurations. In both cases, we observe a clear signal for $\hbar \Delta_{\mu \mathrm{w}} <|J'|- |J|$ from lower band modes delocalized along the chain. States in the upper band are not observed as the microwave coupling from $|0\rangle$ to these states is very small. Only in the topological configuration do we observe an additional signal localized at the boundaries around zero energy, corresponding to the two edge modes. The finite width of the signal is due to microwave power broadening~\cite{suppl_mat}. In Fig.~\ref{fig3}C, we quantitatively show the localization of edge modes at the boundary  by post-selecting experimental runs where at most one particle was created. We observe that the particle populates significantly only the leftmost and rightmost sites, and their second neighbors, as expected from the sub-lattice symmetry (edge states have support on one of the two sub-chains only) and in a good agreement with a parameter-free calculation (black crosses).

For any finite chain, the left and right edge modes hybridize to form symmetric and anti-symmetric states with an energy difference $E_{\rm hyb} \propto J' |J'/J|^N$, which breaks the degeneracy of the edge modes but decreases exponentially with the chain length $N$. While this remains negligible compared to our experimental time scale for a long chain of 14 sites ($E_{\rm hyb} \simeq h\times 20$~kHz), the hybridization is observable for smaller chains. Notably, it gives rise to a coherent transfer of a particle between the two boundaries without involving the bulk modes, as sketched in Fig.~\ref{fig3}D. To observe this, we prepare a particle on the leftmost site using a combination of an addressing beam and microwaves sweeps~\cite{deLeseleuc2017}, and then let the system evolve freely. We show in Fig.~\ref{fig3}E the experimental results for three chains of 4, 6 and 8 sites. The energy $E_{\rm hyb}$ is determined from the frequency of transfer and exhibits the expected exponential scaling, see Fig.~\ref{fig3}F, in excellent agreement with theoretical calculations including the full hopping matrix $J_{ij}$.

\begin{figure*}[htb]
	\includegraphics[width=\textwidth]{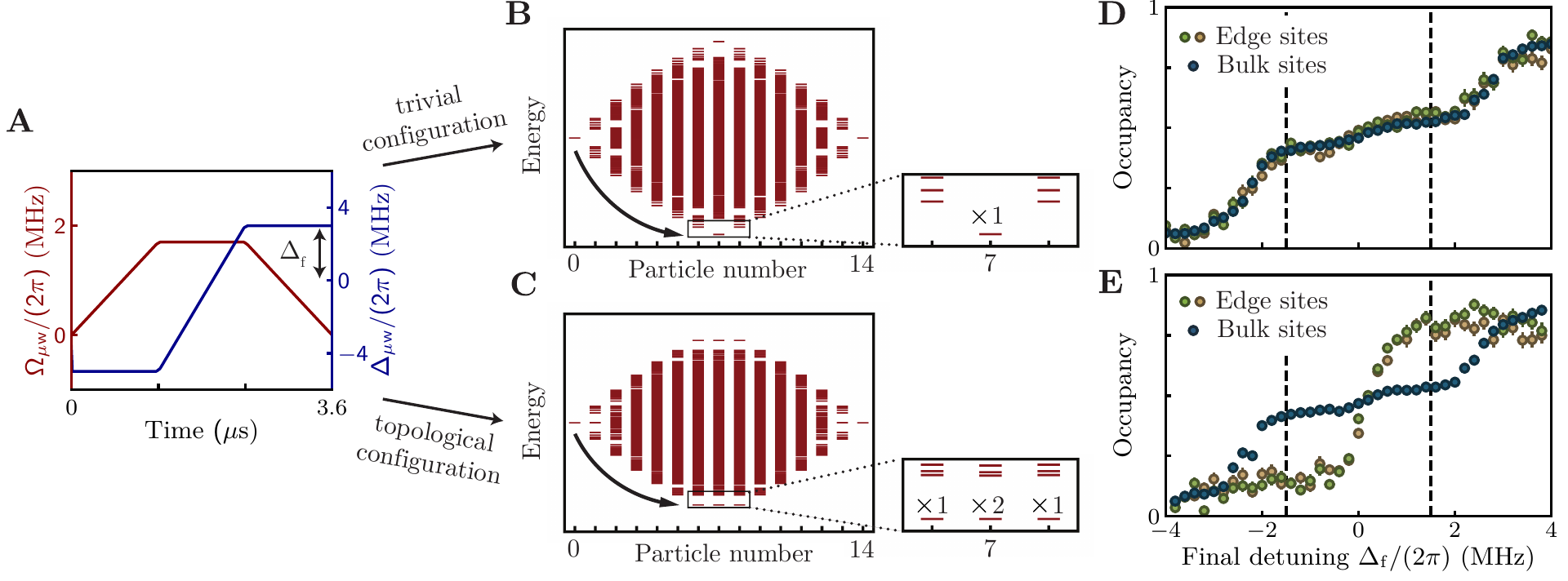}
	\caption{{\bf Preparing the many-body phase at half-filling.} (A) Microwave sweep with time-varying Rabi frequency $\Omega_{\mu \mathrm{w}}$ and detuning $\Delta_{\mu \mathrm{w}}$; the latter ends at $\Delta_{\mathrm{f}}$. (B,C) Energy spectrum of the many-body system in the trivial (top) and the topological (bottom) configuration for different particle numbers. The trivial chain exhibits a single gapped ground state with 7 particles, while	the topological configuration exhibits a four-fold degeneracy involving 6, 7 (two-fold degenerate), and 8 particles.	Starting from the empty chain, the microwave adiabatic sweep loads hard-core bosons in the lattice and prepares the lowest energy states. (D,E) We measure the occupancy of bulk (blue) and edge sites (green and brown) as a function of the final detuning $\Delta_{\mathrm{f}}$. For a sweep ending in the single-particle gap (dashed lines), the bulk of the chain is half-filled. Bosons are loaded in the edge sites of the topological configuration when $\Delta_{\mathrm{f}} > 0$. The error bars represent the standard errors of the mean (s.e.m).
	 \label{fig4_part1}}
\end{figure*}

\section*{Many-body ground state}

We now turn to the study of the many-body system.  In Fig.~\ref{fig4_part1}B-C, we show the full energy spectrum in the trivial and topological configurations, calculated using exact diagonalization, and ordered by increasing number of particles. In the trivial case, there is a single ground state at half-filling. In contrast, the topological configuration exhibits four degenerate ground states corresponding to the bulk half-filled, and which are characterized by additional or missing particles mainly residing at the edges. In order to prepare the ground state, we perform a microwave adiabatic sweep, shown in Fig.~\ref{fig4_part1}A, where the final detuning $\Delta_{\mathrm{f}}$ plays the role of a chemical potential tuning the number of particles loaded in the chain. From a theoretical analysis simulating the full time-evolution, we expect that our ramping procedure ending at a final detuning $|\hbar \Delta_{\mathrm{f}}| < |J|-|J'|$ prepares the ground state with high fidelity~\cite{suppl_mat}.

We present in Fig.~\ref{fig4_part1}D-E the dependence of the local density of particles on $\Delta_{\mathrm{f}}$: the bulk sites occupancy (blue curves) exhibits a characteristic plateau at half-filling within the single particle gap. Especially, the fluctuations of the number of particles in the bulk are strongly reduced with a probability of 48~\% to find exactly 6 particles on the 12 bulk sites (mainly decreased from 100~\% by detection errors~\cite{suppl_mat}). While the local bulk properties are independent of the topology of the setup, the situation is drastically different for the edge occupancy: in the trivial configuration, the edge sites behave as the bulk sites, whereas for the topological chain the boundaries remain depleted for $\Delta_{\mathrm{f}}<0$ and exhibit a sharp transition to full occupancy for $\Delta_{\mathrm{f}}>0$. This behavior is consistent with the expected ground state degeneracy.

We gain more insight about the many-body state by analyzing the correlations between particles, that we can measure as our detection scheme provides the full site-resolved particle distribution. In the strongly dimerized regime $|J| \gg |J'|$, we expect the $\sim N/2$ particles in the bulk to be highly correlated as they can minimize their energy by each delocalizing on a dimer (two sites connected by a strong link $J$). The picture remains valid even in our regime where $|J |\simeq 2.6 |J'|$.  We measure a large and negative density-density correlation $C^z(2i,2i+1) =\langle  Z_{2i} Z_{2 i +1}\rangle \simeq -0.67(1) $ with $Z_{i}= 1-2 b_{i}^\dag b_{i}^{}$, corresponding to a suppressed probability to find two particles on the same dimer. We also access the off-diagonal correlations, $C^x(i,j) = \langle X_{i} X_{j}\rangle$ with $X_{i}=b_{i}+b_{i}^{\dag}$ measuring the coherence between two sites $i$ and $j$, by applying a strong microwave pulse before the detection which rotates the local measurement basis around the Bloch sphere. We obtain $C^x(2i,2i+1) \simeq +0.48(2)$ indicating that a particle is coherently and symmetrically delocalized on two sites forming a dimer. Furthermore, our detection scheme allows us to determine string order parameters, which have emerged as an indicator of topological states~\cite{Kennedy92,Hida92}:
\begin{equation}
 C^{z}_{\rs{string}}=- \left \langle    Z_{2}\:   e^{i \frac{\pi}{2} \sum_{k=3}^{N-2} Z_{k}} \: Z_{N-1} \right\rangle 
\end{equation}
and in analogy for $ C^{x}_{\rs{string}}$. Indeed, we measure a finite string order in the topological phase with $C^{z}_{\rs{string}}= 0.11(2)$
and $C^{x}_{\rs{string}}= 0.05(2)$,  while in the trivial phase they are consistent with zero, e.g., $C^{z}_{\rs{string}}= - 0.02(3)$. All measured correlators are in good agreement with simulations~\cite{suppl_mat}.

\begin{figure*}[htb]
	\includegraphics[width=0.8\textwidth]{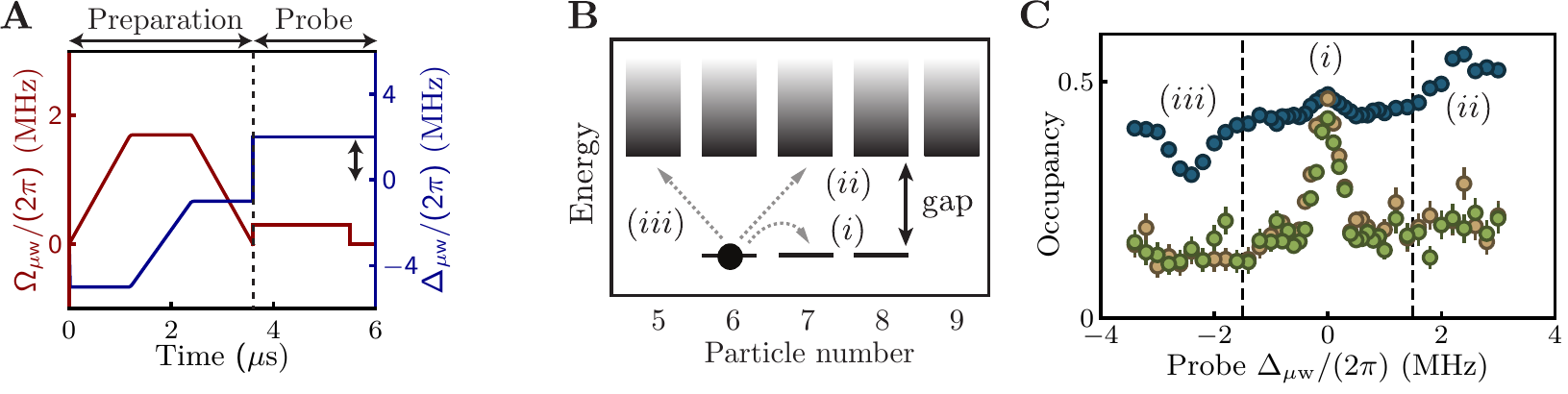}
	\caption{{\bf Probing the SPT phase degeneracy and bulk excitation gap.}
		(A) A microwave sweep ending at $\Delta_{\mathrm{f}}/(2\pi) = -1$~MHz first prepares the many-body ground state with 6 particles, and we then apply for $2 \, \mu$s a microwave probe with a Rabi frequency $\Omega_{\mu \mathrm{w}}/(2\pi) = 0.3$~MHz and a variable detuning $\Delta_{\mu \mathrm{w}}$. (B) Zoom on the bottom of the energy spectrum of a chain in the topological configuration. Starting from the ground state with $6$ particles (solid disk), we can (i) reach one of the other degenerate ground states by adding a particle at the edge for zero energy cost. In addition, we can probe the bulk excitation gap by (ii) adding a particle to, or (iii) removing a particle from, the bulk. (C) Measured occupancy of bulk (blue) and edge sites (green and brown) showing the three expected transitions. Error bars are s.e.m.
		\label{fig4_part2}}
\end{figure*}

We now demonstrate the degeneracy of the many-body ground state in the topological phase and the bulk excitation gap. We first prepare the many-body ground state with the bulk at half-filling but empty edge states by an adiabatic sweep ending at $\Delta_{\mathrm{f}}/(2 \pi) = -1$~MHz. We then apply a weak microwave probe at various detunings $\Delta_{\mu  \mathrm{w}}$ (see Fig.~\ref{fig4_part2}A) and observe when particles are created or annihilated in order to probe the excitation spectrum of the many-body ground state. Figure~\ref{fig4_part2}B-C shows the three expected and measured transitions: (i) a particle is added at zero energy at the edge, and we reach another of the four degenerate ground states, (ii) particles are added to the bulk, which requires at least the bulk gap in energy, while (iii) particles are removed from the bulk, which appears as a dip at negative detuning.

\section*{Probing the protecting symmetry}

\begin{figure}[htb]
	\includegraphics[width=1\columnwidth]{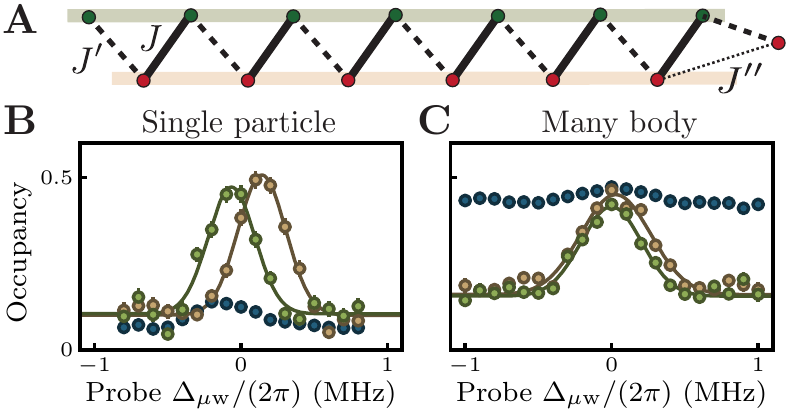}
	\caption{{\bf Perturbation and robustness of the bosonic topological phase.} 
		(A) The rightmost site is shifted upwards to give a finite hopping amplitude $J''$ to the second neighbor. (B,C) Probability to find a particle in the left (green) and right (brown) edge sites when scanning the detuning $\Delta_{ \mu \mathrm{w}}$ of the microwave probe. The experiment is performed either on (B) an initially empty chain to observe the energy difference between the two single-particle edge modes caused by the perturbation $J''$ or (C) on the many-body ground state with a half-filled bulk (6 particles in a 14-site chain) to observe the protection of the ground state degeneracy. Solid lines are Gaussian fits from which we extract an energy difference of $0.21(1)$~MHz in (B) and $0.03(2)$~MHz in (C). 
		\label{fig5}}
\end{figure}

We finally probe the robustness of the four-fold ground state degeneracy to small perturbations, which respect the protecting symmetry $\mathcal{S}_{\rs B}$. To do so, we distort the chain on one side by moving the rightmost site out of the sub-lattice $B$, see Fig.~\ref{fig5}A. As the edge site and its second neighbor are not at the `magic angle' anymore, this creates a coupling $J''/h \simeq 0.26$~MHz between them. This perturbation breaks the chiral symmetry protecting the fermionic SSH model, and correspondingly leads to a splitting of the single-particle edge modes. However, such a perturbation commutes with the symmetry $\mathcal{S}_{\rs B}$ and therefore should not break the many-body ground state degeneracy. To check these expectations, we first repeat the spectroscopic measurement in the single-particle regime (applying the microwave probe on an empty chain, as shown in Fig.\ref{fig3}A), and observe a splitting of the edge modes, see Fig.~\ref{fig5}B. In contrast, the spectroscopic measurement for the bosonic many-body ground state (applying the probe after the adiabatic preparation reaching half-filling of the bulk, as done in Fig.~\ref{fig4_part2}) indeed reveals a degenerate ground state, see Fig.~\ref{fig5}C. In~\cite{suppl_mat}, we checked that when we prepare the ground state with a half-filled bulk, i.e., when $\Delta_{\mathrm{f}}$ lies in the region $|\hbar \Delta_{\mathrm{f}}| < |J|-|J'|$, the spectroscopic measurement reveals a symmetry protected ground state degeneracy. 

The above experiment illustrates that, in contrast to a non-interacting SPT phase, the robustness of the bosonic many-body ground state at half-filling cannot be understood at the single-particle level. To gain an intuition for the differences between the SPT phase of non-interacting fermions and of hard-core bosons, we use the following simple picture. Considering only the three rightmost sites (the edge and a dimer), and taking the perturbative limit ($J \gg J',J''$), we first obtain the energy of having no particle on the edge site and one delocalized on the dimer: $-J -(J'+J'')^2/\left(2J\right)$ (the second term is an energy correction due to virtual hopping of the particle from the bulk to the edge). On the contrary, when there is one particle on the dimer and one on the edge, we obtain $-J - (J' \pm J'')^2/\left(2J\right)$ with an energy correction now depending on the particle quantum statistics ($+$ sign for bosons, $-$ for fermions, due to commutation rules). More details can be found in S3.3 of~\cite{suppl_mat}. This simplified model captures why the fermionic degeneracy is broken by the $J''$ term, while this is not the case for hard-core bosons. 

\section*{Discussion and Outlook}

In this work, we prepared quantum many-body states in two topologically different phases, and observed four characteristic signatures of a SPT phase for interacting bosons in one-dimension: (i) a ground state degeneracy characterized by zero-energy edge states, (ii) an excitation gap in the bulk, (iii) a non-vanishing string order, and (iv) a robustness of these properties against perturbations respecting the protecting symmetry $\mathcal{S}_{\rs B}$. 

Furthermore, we point out that the symmetry protecting our SPT phase defines the symmetry group $U(1)\times \mathbb{Z}_{2}^{T}$, which can also protect the Haldane phase of an anti-ferromagnetic spin-1 chain~\cite{Haldane1983}. We show numerically in~\cite{suppl_mat} that our quantum many-body ground state is in the same SPT phase as the celebrated Haldane state, as they can be smoothly connected by adding hopping terms with complex amplitudes as well as additional Ising type interactions $V_{ij} \propto Z_{i}Z_{j}$, both terms respecting $\mathcal{S}_{\rs B}$. 

Our work demonstrates that Rydberg platforms, which combine flexible geometries and a natural access to the strongly correlated regime via the hard-core constraint, are capable to explore unconventional quantum many-body states of matter. Combining ideas for obtaining topological band structures in higher dimensions~\cite{Weber2018}, it opens the prospect for the realization of long-range entangled states such as fractional Chern insulators with anyonic excitations.

\section*{Acknowledgment}
This work benefited from financial support by the EU (FET-Flag 817482, PASQUANS), and  ``Investissements d'Avenir'' LabEx PALM (ANR-10-LABX-0039-PALM, projects QUANTICA and XYLOS), H.P.~B is supported by the European Union under the ERC consolidator grant SIRPOL (grant N. 681208).

\bibliography{ssh_refs}


\end{document}


\title{Supporting Online Material for:\\ Experimental realization of a symmetry protected topological phase \\ of interacting bosons with Rydberg atoms}

	\author{Sylvain de L\'es\'eleuc}
	\email[The four authors contributed equally to this work.]{}
	\affiliation{Laboratoire Charles Fabry, Institut d'Optique Graduate School, CNRS, Universit\'e Paris-Saclay, 91127 Palaiseau Cedex, France}
	\author{Vincent Lienhard}
	\email[The four authors contributed equally to this work.]{}
	\affiliation{Laboratoire Charles Fabry, Institut d'Optique Graduate School, CNRS, Universit\'e Paris-Saclay, 91127 Palaiseau Cedex, France}
	\author{Pascal Scholl}
	\affiliation{Laboratoire Charles Fabry, Institut d'Optique Graduate School, CNRS, Universit\'e Paris-Saclay, 91127 Palaiseau Cedex, France}
	\author{ Daniel Barredo}
	\affiliation{Laboratoire Charles Fabry, Institut d'Optique Graduate School, CNRS, Universit\'e Paris-Saclay, 91127 Palaiseau Cedex, France}
	\author{Sebastian~Weber }
	\email[The four authors contributed equally to this work.]{}
	\affiliation{Institute for Theoretical Physics III and Center for Integrated Quantum Science and Technology, University of Stuttgart, 70550 Stuttgart, Germany}
	\author{Nicolai Lang }
	\email[The four authors contributed equally to this work.]{}
	\affiliation{Institute for Theoretical Physics III and Center for Integrated Quantum Science and Technology, University of Stuttgart, 70550 Stuttgart, Germany}
	\author{Hans Peter B\"uchler}
	\affiliation{Institute for Theoretical Physics III and Center for Integrated Quantum Science and Technology, University of Stuttgart, 70550 Stuttgart, Germany}
	\author{Thierry Lahaye}
	\affiliation{Laboratoire Charles Fabry, Institut d'Optique Graduate School, CNRS, Universit\'e Paris-Saclay, 91127 Palaiseau Cedex, France}
	\author{Antoine Browaeys}
	\affiliation{Laboratoire Charles Fabry, Institut d'Optique Graduate School, CNRS, Universit\'e Paris-Saclay, 91127 Palaiseau Cedex, France}
	\date{\today}

\maketitle

\clearpage
\newpage

\twocolumngrid

\renewcommand{\thefigure}{S\arabic{figure}}
\renewcommand{\thetable}{S\arabic{table}}
\renewcommand{\theequation}{S\arabic{equation}}

\renewcommand\thesection{S\arabic{section}}
\renewcommand\thesubsection{S\arabic{section}.\arabic{subsection}}
\renewcommand\thesubsubsection{S\arabic{section}.\arabic{subsection}.\arabic{subsubsection}}

\makeatletter
\def\p@subsection{}
\def\p@subsubsection{}
\makeatother

\setcounter{figure}{0}

%

\makeatletter
\makeatletter \renewcommand{\fnum@figure}
{\figurename~\thefigure}
\makeatother

\makeatletter
\makeatletter \renewcommand{\fnum@table}
{\tablename~\thetable}
\makeatother

\tableofcontents

In this supplement, we first describe our experimental methods and compare the measurements with simulations (S1). We then give a more complete theoretical analysis of the SSH model, comparing the case of fermions (S2) to hard-core bosons (S3), and finally connecting the topological phase of the bosonic SSH chain to the Haldane phase (S4).
 
\section{Experimental methods and comparison to simulations}

We compare the experimental results discussed in the main text with simulations with no adjustable parameters. For better comparison between experiment and theory, the simulations take into account experimentally estimated preparation and detection errors. 

\subsection{Preparation and detection errors}

\begin{figure}
	\centering
	\includegraphics[width= 1\columnwidth]{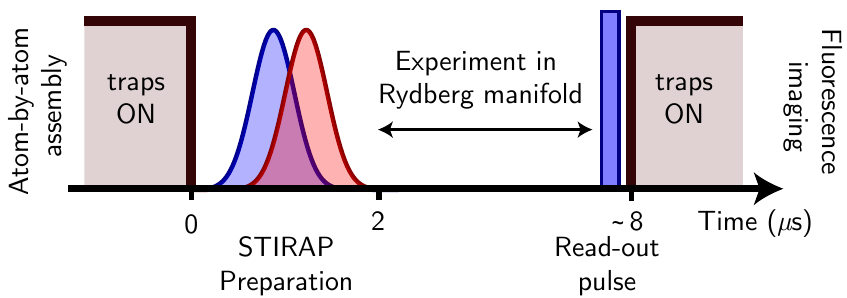}
	\caption{Typical experimental sequence. After assembling atom-by-atom a chain of ground state atoms, we transfer them to the Rydberg $60S_{1/2}$ level with a two-photon STIRAP pulse lasting $2 \, \mu$s. We then perform an experiment in the Rydberg manifold (e.g., microwave spectroscopy, sweep...), which is ended by a $0.3 \, \mu$s read-out pulse depumping atoms in the $60S_{1/2}$ state back to the electronic ground states. These atoms are recaptured by the tweezers and detected in the fluorescence image, while atoms in the $60P_{1/2}$ state are lost.} 
	\label{fig:experimental_sequence}
\end{figure}

Our experimental sequence, sketched in Fig.~\ref{fig:experimental_sequence}, starts with the preparation of an array of $^{87}$Rb atoms, each trapped in an individual tweezers, using the atom-by-atom assembly technique~\cite{Barredo2016}. The atoms are then transferred from the electronic ground-state level $\ket{5S_{1/2},F=2,m_F=2}$ to the Rydberg s-level $\ket{60S_{1/2},m_J=1/2}$ using a two-photon STIRAP (STImulated Raman Adiabatic Passage) pulse~\cite{Cubel2005,Deiglmayr2006,review_stirap,Higgins2017}, detailed in Ref.~\cite{deLeseleucThesis}, pp.~140--147. There is a finite preparation error, measured to be $\eta = 5-7$~\%, that an atom is not transferred to the Rydberg level. This gives rise to lattice defects, that are taken into account in the numerical simulations by averaging over typically $\sim 1000$ lattice realizations.

For read-out, we de-excite the atoms in the $60S_{1/2}$ level to the electronic ground state by shining during $0.3 \, \mu$s a 475~nm beam resonant with the short-lived $5P_{1/2}$ state. Atoms in the $60P_{1/2}$ state are not affected by this pulse. Then, the de-excited atoms are recaptured in the tweezers with a probability $1-\varepsilon$, while atoms left in the Rydberg manifold are lost with a probability $1-\varepsilon'$~\cite{deLeseleuc2018b}. We estimate experimentally that the detection errors are $\varepsilon = 0.05(1)$ and $\varepsilon' =0.05(1)$. In the simulations, they are taken into account by a Monte Carlo sampling of the numerical results. 



\subsection{Single-particle spectra} \label{sec:comparison_spectra}

\begin{figure}
	\centering
	\includegraphics[width= 1\columnwidth]{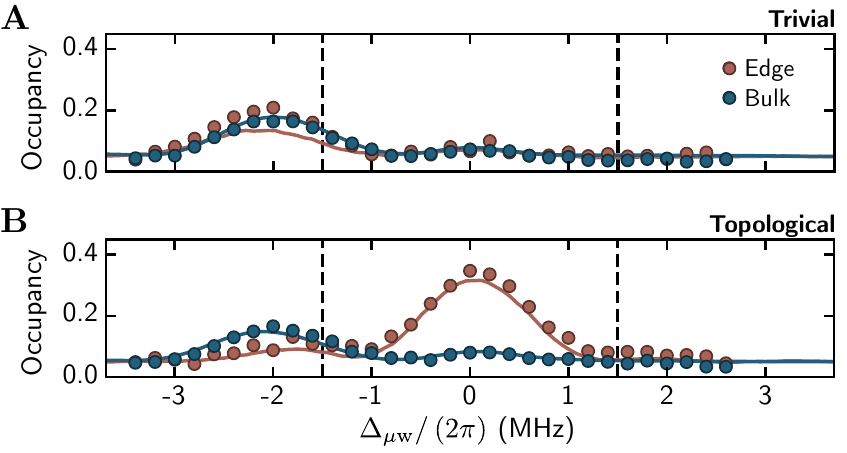}
	\caption{Comparison of the measured single-particle spectra (disks), taken from Fig. 2B of the main text, with theory (lines). The spectra show the probability to find a particle on the left or right site of the SSH chain, as well as the site-averaged probability to find a particle in the bulk as a function of the microwave detuning $\Delta_{\mu\mathrm{w}}$. The dashed lines symbolize the band gap. The measured and calculated spectra match for the trivial configuration~(A) as well as for the topological configuration~(B). Error bars (s.e.m) are smaller than the symbol size.} 
	\label{fig:comparison_spectra}
\end{figure}

We first reproduce numerically the microwave spectroscopy experiment starting from an empty chain, probing the single-particle spectrum of a trivial and topological chain. For the simulation, the Hamiltonian consists of the bosonic SSH model [see Eq.~(1) of the main text] and the interaction with the microwave probe with Rabi frequency $\Omega_{\mu\mathrm{w}}/2\pi = \unit[0.2]{MHz}$, which we treat in the rotating wave approximation. We use a Krylov subspace method to compute the time evolution of the system. After an evolution time of $\unit[0.75]{\mu s}$, we calculate the probability to find a particle on a given lattice site, see Fig. \ref{fig:comparison_spectra}. The simulation, without any adjustable parameters, reproduces very well the experimental data shown in Fig. 2B of the main text, including the positions and widths (due to microwave power broadening) of the spectroscopic features. 


\subsection{Hybridization of edge modes}

As discussed in the main text, the energy of the symmetric and antisymmetric edge modes differs by the hybridization energy $E_\text{hyb}$. Here, we give details on Fig. 2F of the main text, which illustrates the scaling of the hybridization energy with the system size, and compare Fig. 2E of the main text with theory.

\begin{figure}[h]
	\centering
	\includegraphics[width= 1\columnwidth]{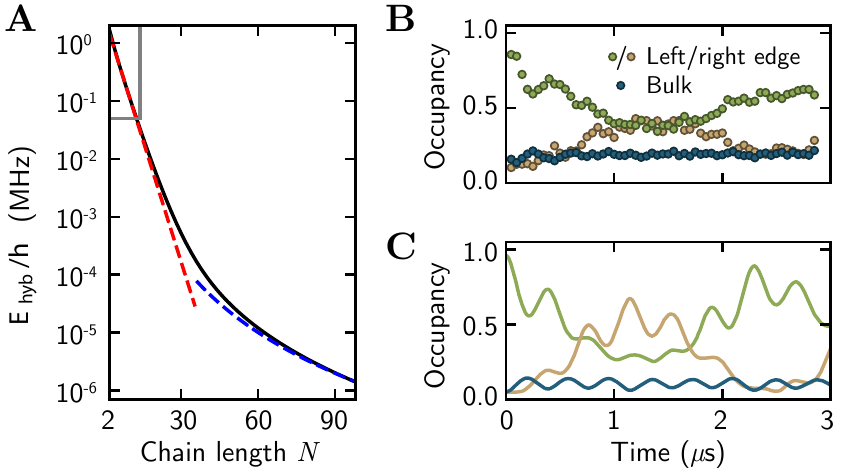}
	\caption{(A) Theoretical scaling of the hybridization energy $E_\text{hyb}$. The red curve is an exponential fit for short chain lengths. For longer chains, the hybridization energy scales algebraically as $1/N^4$ (blue curve). The region surrounded by gray lines corresponds to the range of chain lengths shown in Fig. 2F of the main text. (B) Measured and (C) simulated time evolution of the probability to find a particle on the left or right edge or in the bulk after creating a localized particle on the left edge of a topological chain of 6 sites.} 
	\label{fig:simulation_hybridization}
\end{figure}

We obtain the hybridization energy by diagonalizing the coupling matrix $J_{ij}$ for different chain lengths up to $N = 100$~sites, see Fig.~\ref{fig:simulation_hybridization}A. After initially decreasing exponentially, $E_\text{hyb}$ scales algebraically with the chain length, as the direct coupling $J_{1,N} \propto 1/N^4$ between the edges dominates over the higher-order coupling via nearest neighbor interactions $J_{i,i+1}$. The $1/N^4$ scaling is a combination of the $1/R^3$-dependence of the dipolar interaction and the pair of edge sites getting closer to the `magic angle'. Note that the transition to the algebraic regime happens for significantly longer chains than studied experimentally.	

As discussed in the main text, the hybridization energy is determined experimentally by measuring the frequency of the particle transfer between the two edges. For this, a localized particle is created on the left edge using a combination of an addressing beam and microwave sweeps (see Ref.~\cite{deLeseleucThesis}, p.155), with an efficiency of $\sim 94$~\%. We then observe the dynamics of this particle. Figure~\ref{fig:simulation_hybridization}B,C compares measurements, taken from Fig.~2E of the main text, with a simulation for a chain of 6 atoms. The dominant oscillation frequency, which gives the hybridization energy, agrees well between theory and experiment. The high-frequency oscillations which are visible in the simulation average out in the experiment due to shot-to shot fluctuations of atomic positions as well as motions of atoms, causing varying hopping strengths. The transfer probability is smaller than one because of lattice defects, coming from the imperfect preparation in the Rydberg state. 


\subsection{Preparation of the many-body ground state}

The experiments conducted in the many-body regime rely on the preparation of the ground state of the bosonic SSH model with a microwave sweep. In presence of a microwave drive of Rabi frequency $\Omega_{\mu\mathrm{w}}$ and detuning $\Delta_{\mu \mathrm{w}}$, the Hamiltonian reads:
\begin{equation}
H = H_{\rm SSH} + \frac{\hbar \Omega_{\mu \textrm{w}}}{2} \sum_i \left[ b_i^\dag + b_i^{} \right] - \hbar \Delta_{\mu \textrm{w}} \sum_i b^\dag_{i} b_i^{}\, ,
\label{eq:ssh_microwave}
\end{equation}
where $H_{\rm SSH}$ is given by Eq.~(1) of the main text. 

\begin{figure}
	\centering
	\includegraphics[width= 1\columnwidth]{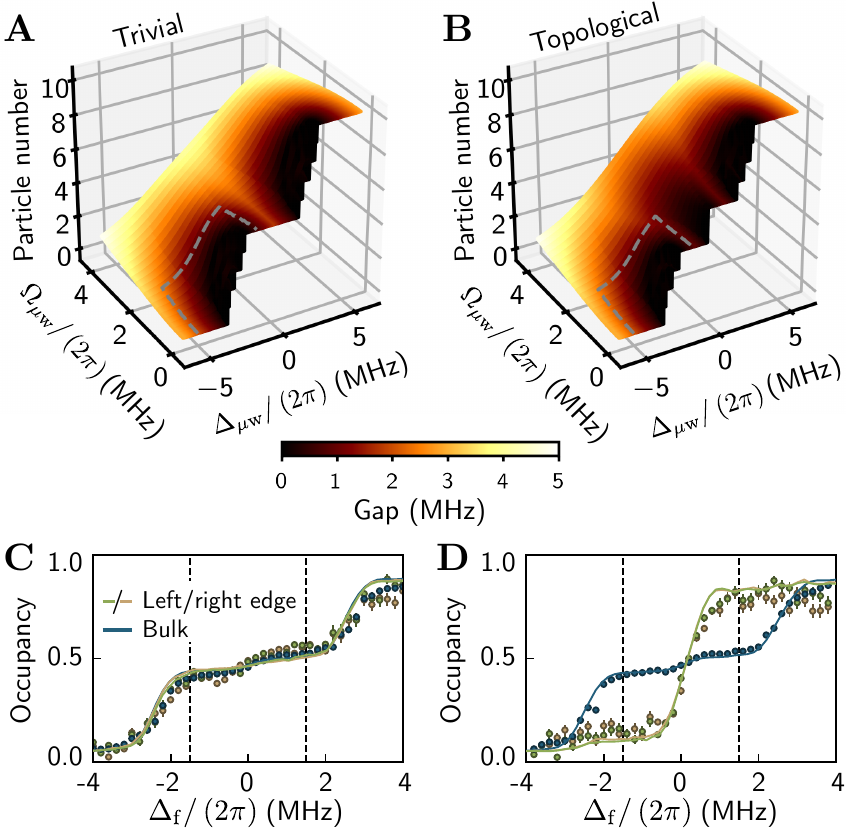}
	\caption{Number of particles in the ground state of the trivial (A) and topological (B) SSH chain of 10 sites in presence of a microwave drive of Rabi frequency $\Omega_{\mu \mathrm{w}}$ and detuning $\Delta_{\mu \mathrm{w}}$ [see Eq.~(\ref{eq:ssh_microwave})]. The colormap indicates the gap between the ground state and the first excited state. (C,D) Measured (disks) and simulated (lines) occupancy of the bulk and edge sites after a microwave sweep, represented as a gray dashed line in (A,B), ending at a varying detuning $\Delta_{\mathrm{f}}$.}
	\label{fig:adiabatic_sweep}
\end{figure}

Figure~\ref{fig:adiabatic_sweep}A-B shows the results of a numerical calculation of the number of particles in the ground state of Eq.~\eqref{eq:ssh_microwave} as a function of the microwave parameters, for a chain of $N = 10$ sites in the trivial and topological configurations. The detuning $\Delta_{\mu \mathrm{w}}$ acts as a chemical potential: for large negative (positive) values, the chain is completely empty (filled), while there is a finite region for a detuning around 0 where the bulk of the chain is half-filled. The SPT ground state degeneracy of the topological chain is seen as a jump of the number of particles from $N/2-1$ to $N/2+1$ at $\Delta_{\mu \mathrm{w}} = 0$ and $\Omega_{\mu \mathrm{w}} = 0$, corresponding to the loading of particles in the two edge states. At $\Delta_{\mu \mathrm{w}}/(2\pi) \simeq \pm 2$~MHz the number of particles also exhibits a step-like behavior. These latter two regions would correspond to a metallic phase in the fermionic SSH model, when the chemical potential lies in the lower or upper band of single-particle eigenstates. 

In Fig.~3 of the main text, we used a microwave sweep ending at varying detuning $\Delta_{\mathrm{f}}$ to observe the features described above. Figure~\ref{fig:adiabatic_sweep}C-D compares the experimental results to numerical simulations taking into account preparation and detection errors, showing an excellent agreement. The small jump of the bulk density around $\Delta_{\mathrm{f}} = 0$, even in the trivial configuration, is caused by preparation errors creating lattice defects: such a defect gives rise to two chains, one of which starts with a weak link, and thus supports a zero-energy edge-state which gets populated when $\Delta_{\rm f}$ becomes positive. The step-like behavior of the number of particles seen in the calculations at $\Omega_{\mu \mathrm{w}} = 0$ around $\Delta_{\mathrm{f}}/(2\pi) \simeq \pm 2$~MHz is not observed experimentally, due to the vanishing gap between the final ground state and the first excited state.

In contrast, in the half-filled region, we observe a finite gap and the system corresponds to an insulator. In the topological configuration, the gap closing at $\Delta_{\mu \mathrm{w}} = 0$ is due to the ground state degeneracy caused by zero-energy edge states. Additionally, we remark the existence of a path in the parameter space $(\Omega_{\mu \mathrm{w}},\Delta_{\mu \mathrm{w}})$ connecting the empty chain to the half-filled state, where the gap never closes (even in the limit of an infinite number of particles), see the dashed line in Fig.~\ref{fig:adiabatic_sweep}B. This enables the adiabatic preparation of the many-body ground state at half-filling with empty edges (using a sweep ending at $\Delta_{\mathrm{f}}/(2\pi) \simeq -1$~MHz) or filled edges ($\Delta_{\mathrm{f}}/(2 \pi) \simeq 1$~MHz), which we demonstrate.

\begin{figure}
	\centering
	\includegraphics[width= 1\columnwidth]{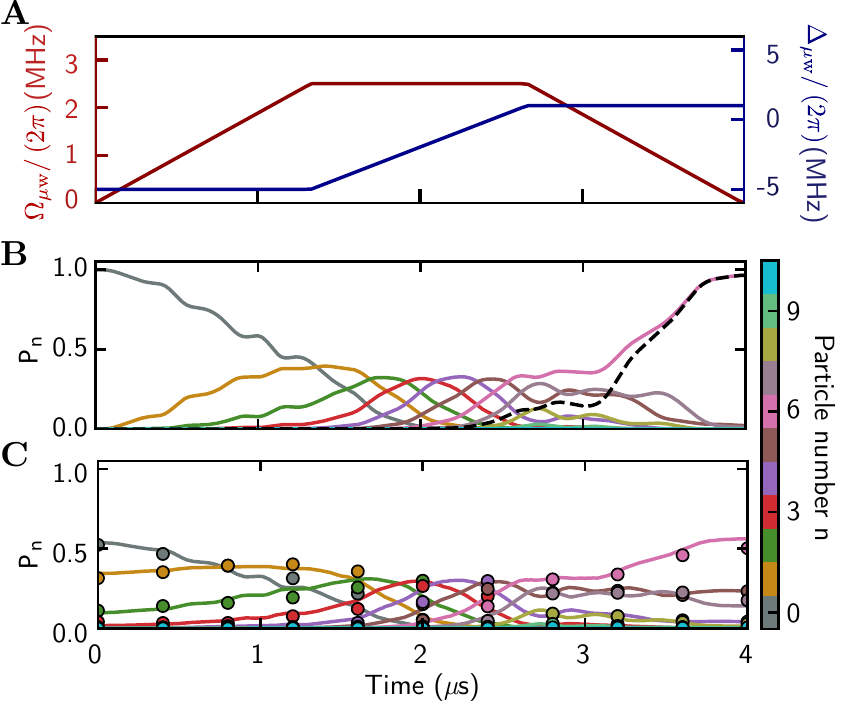}
	\caption{(A) Microwave sweep ending at $\Delta_{\mathrm{f}}/2\pi = \unit[+1]{MHz}$, used to prepare the ground state of the topological setup with filled edge states (here, for a chain of 10 sites). (B)~Numerically calculated evolution of the number of excitations during the sweep, neglecting
		preparation and detection errors. The probability $P_n$ for finding $n$ excitations within the system is depicted. As expected, there is mainly 6 particles at the end of the sweep. The dashed curve shows the overlap with the target state with a final value of $0.965$. (C) Evolution of the number of excitations, measured experimentally (disks) and calculated (lines) including preparation and detection errors  $\varepsilon = 0.06$ and $\varepsilon' = 0.07$, slightly higher for this dataset.} 
	\label{fig:video_sweep}
\end{figure}

To do so, we calculate the fidelity of the ground state preparation by simulating the time evolution of the system under the adiabatic sweep ending at $\Delta_{\mathrm{f}}/(2 \pi) = 1$~MHz, shown in Fig.~\ref{fig:video_sweep}A. Panel~B shows how the overlap with the targeted many-body ground state develops during the sweep, reaching its final value $0.963$ (neglecting preparation and detection errors). In addition, the figure reveals how the number of particles in the system evolves. When including the experimental errors, the simulation is in very good agreement with the data, see Fig.\ref{fig:video_sweep}C.

\subsection{Correlations and string orders}

\begin{figure}
	\centering
	\includegraphics[width= 1\columnwidth]{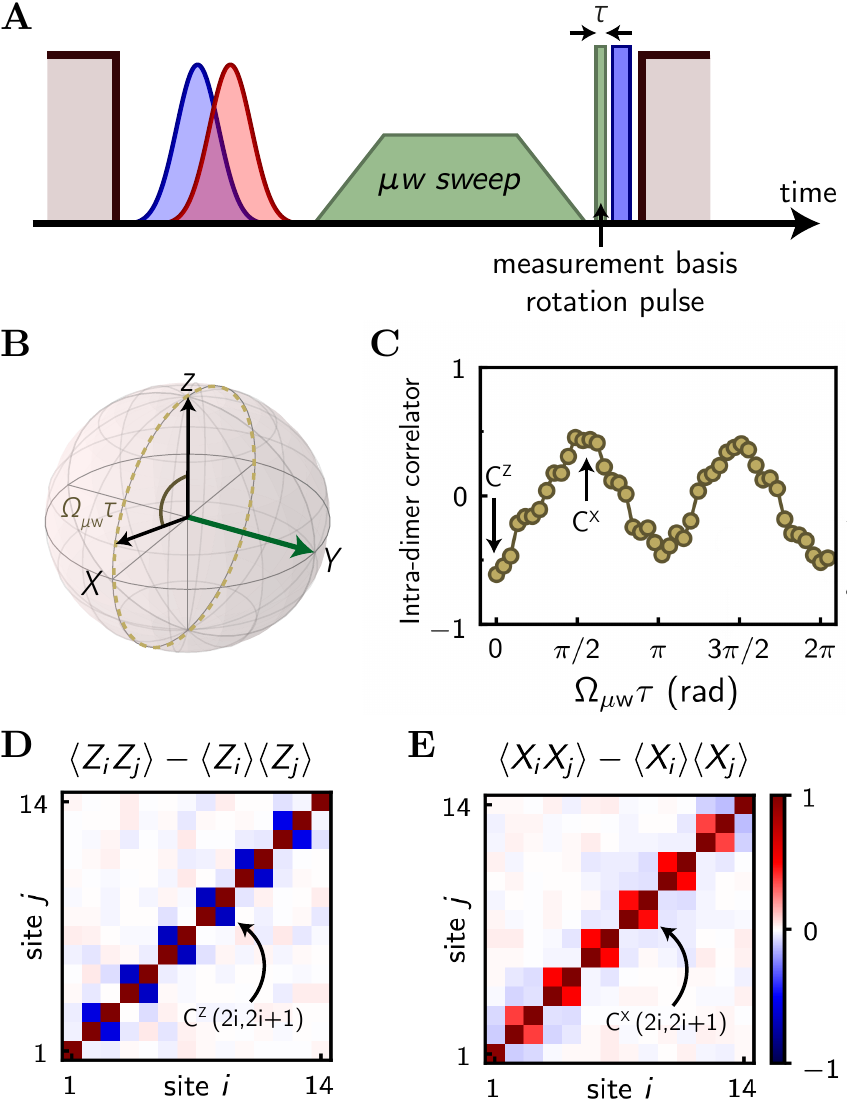}
	\caption{(A) A microwave sweep first prepares the half-filled ground state of a topological chain of 14 sites. Before shining the read-out pulse, we apply  a strong microwave field $\Omega_{\mu \mathrm{w}}/(2 \pi) \simeq 14$~MHz during a time $\tau$ to rotate the measurement basis, as shown in the Bloch sphere representation (B). (C) Measured intra-dimer correlator as a function of the pulse area $\Omega_{\mu \mathrm{w}} \tau$. (D,E) Full correlation maps for two sites $i$ and $j$ in the chain obtained when measuring the $Z$ (D, $\Omega_{\mu \mathrm{w}}\tau = 0$) and $X$ (E, $\Omega_{\mu \mathrm{w}} \tau = \pi/2$) observables.} 
	\label{fig:correlations}
\end{figure}

In the main text, we measured the correlations $C^{z,x}$ and string order parameters $C^{z,x}_\text{string}$  of the many-body ground state. They were obtained for two observables $Z_i = 1-2b_i^\dag b_i^{}$ and $X_i = b_i^{} + b_i^{\dag}$. Here, we first explain how we measured them and then compare the measured $C^{z,x}$ and  $C^{z,x}_\text{string}$ to numerical simulations.

Figure~\ref{fig:correlations}A shows how we perform the experiments. After a microwave sweep preparing the half-filled ground state, we apply a strong microwave pulse that rotates the measurement basis along the $X-Z$ plane, as represented in the Bloch sphere picture in Fig.~\ref{fig:correlations}B. We choose a large Rabi frequency $\Omega_{\mu \mathrm{w}}/(2 \pi) = 14$~MHz, much larger than the interaction energies, to minimize their effects during the rotation. The measured correlations between two sites forming a dimer (connected by a strong link $J$) is shown in Fig.~\ref{fig:correlations}C as a function of the pulse area. A pulse lasting $\tau \simeq 17$~ns rotates the measurement basis from $Z$ to $X$. For completeness, we show the full correlation maps in Fig.~\ref{fig:correlations}D-E. As expected (see the discussion in the main text), we recognize strong correlations for two sites connected by a strong link, both for the $Z$ and $X$ observables. Let us note that we observe inter-dimer correlations that are stronger when measuring along the $X$ axis, which is also predicted in numerical calculations.

Table~\ref{tab:correlators_comp} compares the measured correlators to numerical simulations. The agreement is excellent for measurements along the $Z$ axis, whereas $C^x$ and $C^x_\text{string}$ are below the predicted values, suggesting that the rotation of the measurement basis suffers from experimental imperfections.

\begin{table}
	\begin{tabular}{l||c|c|c|c}
		& $C^z$ & $C^x$ & $C^z_\text{string}$  & $C^x_\text{string}$  \\
		\hline
		Th. (no errors) & -0.96 & 0.98 & 0.78 & 0.88\\
		Full simulation & -0.69(1) & 0.68(2) & 0.11(2) & 0.10(2)\\
		Experiments & -0.67(1) & 0.48(2) & 0.11(2) & 0.05(2)\\
	\end{tabular} 
	\caption{
	Theoretical predictions (with and without experimental imperfections) and experimental measurements of the intra-dimer correlators $C^z$ and $C^x$, as well as of the string order parameters $C^z_\text{string}$ and $C^x_\text{string}$.
	}
	\label{tab:correlators_comp}
\end{table}

\subsection{Probing the robustness against perturbations}

\begin{figure}[b]
	\includegraphics[width= 1\columnwidth]{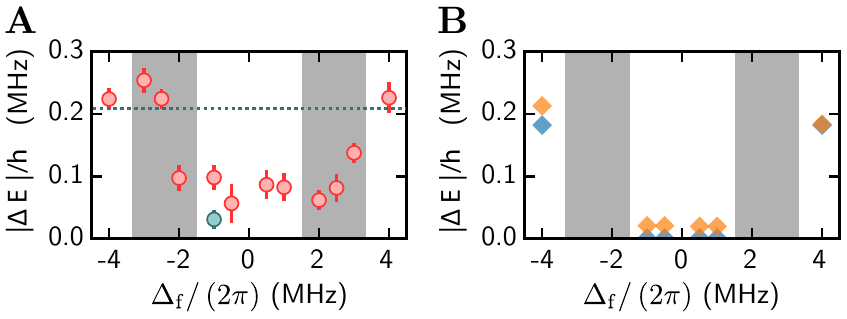}
	\caption{Energy difference $\Delta E$ between the left and right edge states extracted from a spectroscopy experiment performed after a microwave sweep ending at a detuning $\Delta_{\mathrm{f}}$. (A) Experimental values for two datasets. Blue: data presented in the main text, the dashed line corresponding to the single-particle spectroscopy performed on an empty chain. Red: additional data where we vary $\Delta_{\mathrm{f}}$. (B) Numerical simulations including (yellow) or excluding (blue) van der Waals terms.\label{fig:edge_state_splitting}} 
\end{figure}

We present experimental data and numerical simulations completing the study shown in Fig.~5 of the main text, where we break the chiral symmetry by engineering a perturbation introducing a coupling $J''$ between the rightmost site and its second neighbor. Following a microwave sweep ending at a given detuning $\Delta_{\mathrm{f}}$, we perform a spectroscopy of the many-body state corresponding to this final detuning. We extract an energy difference $|\Delta E|$ by fitting with Gaussian functions the measured occupancies of the left and right edge sites. Fig.~\ref{fig:edge_state_splitting}A shows $|\Delta E|$ as a function of $\Delta_{\mathrm{f}}$ for two datasets. The blue dashed line (single particle spectroscopy) and the point at $\Delta_{\mathrm{f}}/(2\pi) = -1$~MHz corresponds to the two measurements presented in Fig.~5 of the main text. The red points correspond to a second dataset for which we varied $\Delta_{\mathrm{f}}$. For $\Delta_{\mathrm{f}}$ largely negative (positive), we prepare an empty (filled) chain and the spectroscopy experiment probes the single-particle eigenmodes, whose degeneracy is broken by the $J''$ perturbation. In turn, for $\Delta_{\mathrm{f}}$ in the half-filled region (see Fig.~\ref{fig:adiabatic_sweep}D), the energy difference is much smaller as the ground state degeneracy is protected by the unbroken symmetry $\mathcal{S}_{\rs B}$. For intermediate values of $\Delta_{\mathrm{f}}$ (gray regions), the many-body state is gapless, which precludes an adiabatic preparation, and we observe a smooth transition between the two regions (empty/full or half-filled chains).  

The larger energy difference observed in the red dataset is most likely caused by an electric field gradient creating an energy difference of $\sim 30-40$~kHz between the leftmost and righmost site due to the Stark effect. This gradient was carefully compensated for the other dataset (blue point), however, even in this case, there is a slight measured energy offset $|\Delta E|/h = 0.03(2)$~MHz. We explain it by a different van der Waals (vdW) interaction of the leftmost and rightmost Rydberg atoms with their neighbors when displacing the rightmost site to engineer the $J''$ term. Including vdW terms in the simulation, we indeed obtain an energy difference larger by $\sim 0.02$~MHz as shown in Fig.~\ref{fig:edge_state_splitting}B.

\section{SSH chain with fermions}
\label{sec:ssh}

\subsection{Many-body Hamiltonian}
\label{subsec:mbh}

The SSH model is a paradigmatic example for a symmetry-protected topological
phase (SPT) of non-interacting \textit{fermions}. It is conveniently described
in a setup with open boundary conditions and an even number of lattice sites
$N$; the Hamiltonian reads
\begin{equation}
H_{\rs F}
=-J'\sum_{i=1}^{L}\,c_{2i-1}^\dag c_{2i}
-J\sum_{i=1}^{L-1}\,c_{2i}^\dag c_{2i+1}+\hc
\label{eq:ssh}
\end{equation}
with $L=N/2$ the number of unit cells and the fermionic creation and
annihilation operators $c^{\dag}_{i}$ and $c_{i}$. The Hamiltonian exhibits two
important symmetries which are relevant for the existence of an SPT: particle
number conservation and the chiral (sublattice) symmetry. Particle number
conservation---an intrinsic symmetry for non-interacting fermionic systems---is
described by a unitary representation of $\group{U}(1)$: $R_\phi=e^{i\phi
	\sum_{i}c^{\dag}_{i}c_{i}}$ with $\com{R_\phi}{H_{\rs F}}=0$ for all
$\phi\in[0,2\pi)$. The chiral symmetry is an antiunitary representation of the
group $\mathbb{Z}_2=\{1,S\}$ with $\com{\S}{H_{\rs F}}=0$ and $\S^2=1$ given by
\begin{equation}
\S=\prod_{i=1}^L(c_{2i-1}^\dag-c_{2 i-1})(c_{2i}^\dag+c_{2 i})\circ K
\label{eq:S}
\end{equation}
where $K$ denotes complex conjugation. It is the latter symmetry that protects
the topological phase for non-interacting fermions.

\subsection{Single-particle picture}
\label{subsec:spp}

The Hamiltonian in Eq.~\eqref{eq:ssh} is quadratic in fermion modes and
therefore can be conveniently expressed as
\begin{align}
H_{\rs F}=\vec\Psi^\dag \hat H\vec\Psi
\label{eq:single_particle}
\end{align}
with pseudo-spinor $\vec\Psi=(c_1,c_2,\dots,c_{N})^T$ and the $N\times
N$ matrix $\hat H$. The matrix $\hat H$ can be diagonalized by a unitary
transformation $U$ so that $\tilde{\vec\Psi}=U\vec\Psi$ defines new fermionic
modes which describe the single-particle eigenmodes of the Hamiltonian
$H_{\rs F}$. The ground state(s) $\ket{\Omega}$ of the many-body theory are
then given by the state(s) where all eigenmodes in $\tilde{\vec\Psi}$ with
negative eigenenergy are occupied, i.e., $\ket{\Omega}$ describes the Fermi
sea of the filled lower band. In particular, all properties of the many-body
ground state(s)---such as the existence of edge modes and the stability of
ground state degeneracies against non-interacting perturbations---derive
directly from the single-particle spectrum.

The next task is to derive the constraints on the matrix $\hat{H}$ imposed by
the symmetries $\S$ and $R_\phi$. First, particle number conservation $R_\phi$
does not restrict $\hat H$ in any way as the parametrization
\eqref{eq:single_particle} satisfies this symmetry per construction.  Second,
the chiral symmetry $\S$ and the condition $\S H_{\rs F}
\S^{-1}\stackrel{!}{=}H_{\rs F}$ imposes the non-trivial constraint
\begin{equation}
\vec S \hat H\vec S^{-1}=U_S\hat HU_S^\dag=-\hat H
\label{eq:slsym}
\end{equation}
on the single-particle Hamiltonian $\hat H$, with $\vec S=U_S$ and the unitary
\begin{equation}
U_S
=\mathds{1}_{L\times L}\otimes
\begin{pmatrix}
1 & 0 \\
0 & -1
\end{pmatrix}\,.
\end{equation}
Note that $\S$ is an \textit{antiunitary} operator on the Fock space but
$\vec{S}$ is realized as \textit{unitary} operation on the single-particle
Hamiltonian $\hat H$. However, $\vec S$ is not an ordinary symmetry of $\hat
H$ as it \textit{anticommutes} with the matrix $\hat H$.

This matrix constraint is one of the three ``generic symmetries'' that give rise
to the classification of free fermion theories known as the \textit{10-fold
	way}~\cite{Ryu2010}. (We refer the reader to Ref.~\cite{Ludwig2015} for a
pedagogical introduction.) According to this classification, $H_{\rs F}$ belongs
to the class \textbf{AIII} which features a $\mathbb{Z}$ topological index in
one dimension. In the topological phase ($|J'|<|J|$), this index is non-zero so
that the 10-fold way establishes the former as SPT protected against
\textit{non-interacting} perturbations that respect the chiral symmetry $\S$.

\subsection{Many-body ground states}
\label{subsec:mbgs}

In the \textit{trivial phase} with $|J'|>|J|$, the ground state describes an
insulator at half filling (i.e., the lowest band is completely filled) with
single-particle excitation gap $|J'|-|J|$. The unique ground state takes a
particularly simple form at the special point $J=0$ and $J'>0$:
\begin{equation}
|\Omega\rangle 
= \prod_{i=1}^{L} \left(\frac{c^{\dag}_{2 i-1}+ c^{\dag}_{2 i }}{\sqrt{2}}\right) |0\rangle\,.
\label{eq:gs1}
\end{equation}

In the \textit{topological phase} with $|J'|<|J|$, the ground state is still
insulating with a single-particle gap $|J|-|J'|$ in the bulk, but exhibits a
four-fold ground state degeneracy (up to exponentially small corrections) due to
the appearance of edge states at the boundary. This can be easily understood at
the special point $J'=0$ and $J>0$ where the four ground states take the form
\begin{equation}
|\Omega_{m,\bar{m}}\rangle 
= \left(c_{1}^\dag\right)^{m}\left(c_{N}^\dag\right)^{\bar{m}} \prod_{i=1}^{L-1} \left(\frac{c^{\dag}_{2 i}+ c^{\dag}_{2 i +1}}{\sqrt{2}}\right) |0\rangle\,;
\label{eq:gs2}
\end{equation}
here, $m, \bar{m} \in \{0,1\}$ label the four ground states and describe
the occupancy of the zero-energy edge modes $c_{1}$ and $c_{N}$.
The two phases are separated by a gapless critical point at $|J|=|J'|$.

\section{SSH chain with hard-core bosons}
\label{sec:sshchcb}

\subsection{Jordan-Wigner transformation of the SSH chain}
\label{subsec:jwtsshc}

We are now interested in the bosonic counterpart of the SSH Hamiltonian
\eqref{eq:ssh}. To this end, we apply the Jordan-Wigner transformation $\JW$
which allows for an exact mapping between fermions and hard-core bosons. This
mapping is based on the observation that on each lattice site for both fermions
and hard-core bosons only two states are allowed: empty and occupied. Therefore
hard-core bosons can also be interpreted as spin-$\frac{1}{2}$ degrees of
freedom $\sigma^\pm_{i}$. However, the operators on different sites anticommute
for fermions and commute for bosons; this difference is accounted for by a
non-local mapping, the Jordan-Wigner transformation
\begin{equation}
\JW(c_j)=\prod_{k=1}^{j-1}\sigma_k^z\cdot\sigma_{j}^+\,,
\quad
\JW(c_j^\dag)=\prod_{k=1}^{j-1}\sigma_k^z\cdot\sigma_{j}^-\,,
\label{eq:jw2}
\end{equation}
which implements anticommutation relations on different sites,
$\acom{\JW(c_j)}{\JW(c_i^\dag)} =0$ for $i \neq j$.

\subsubsection{Transformation of the SSH Hamiltonian}
\label{subsubsec:tsshh}

If we apply $\JW$ to the SSH Hamiltonian \eqref{eq:ssh}, we find the
corresponding Hamiltonian $H_{\rs B}\equiv \JW(H_{\rs{F}})$ for hard-core
bosons,
\begin{subequations}
	\label{eq:sshhb}
	\begin{align}
	H_{\rs B}
	&=-J'\sum_{i=1}^L \sigma_{2i-1}^-\sigma_{2i}^+ - J\sum_{i=1}^{L-1} \sigma_{2i}^-\sigma_{2i+1}^+ +\hc \\
	&=-\sum_{i\in A,j\in B}\,J_{ij}\left[b_i^\dag b_j+b_j^\dag b_i\right]\,,
	\end{align}
\end{subequations}
which is an alternating spin-exchange model ($XY$-model). How this connects to
\textit{isotropic} spin chains (in particular, the Haldane chain) is explained
in Section \ref{sec:isotropic}. Note that here open boundary conditions are
crucial because the non-local Jordan-Wigner string makes $\JW(H_f)$ a non-local
bosonic Hamiltonian for periodic boundaries. The Hamiltonian \eqref{eq:sshhb} is
the idealization of what is realized in the experiment if the two Rydberg levels
are identified as the two spin states (and if long-range hopping is ignored).

\subsubsection{Transformation of the symmetries}
\label{subsubsec:ts}

Just as the Hamiltonian, also its (many-body) symmetries can be translated
with $\JW$ into symmetries of the bosonic version. Since $\JW$ is an algebra
isomorphism, commutation relations survive and the Jordan-Wigner transformed
operators are again symmetries of the Hamiltonian $H_{\rs B}$. Indeed,
for \textit{particle number conservation} we find the representation of
$\group{U}(1)$
\begin{equation}
\JW(R_\phi)
=\exp\left[i\phi\sum_i\frac{\mathds{1}-\sigma_i^z}{2}\right]\,,
\end{equation}
with $\com{\JW(R_\phi)}{H_{\rs B}}=0$ for all $\phi\in[0,2\pi)$. Furthermore,
the \textit{chiral symmetry} yields the antiunitary operator
\begin{equation}
\S_{\rs B} \equiv    
\JW(\S)=\prod_{i}^N(b_i+b_i^\dag)\,K
\label{eq:SB}
\end{equation}
with $\com{\S_{\rs B}}{H_{\rs B}}=0$. In the picture of hard-core bosons, the
operator $\S_{\rs B}$ converts particles into holes and vice versa. Note that
the Jordan-Wigner transformation removes the signs in \eqref{eq:S} to end up
with \eqref{eq:SB}.

\subsubsection{Transformation of the many-body ground states}
\label{subsubsec:tmbgs}

If one identifies the fermionic Fock state vacuum $\ket{0}$ with
$\ket{\Uparrow}=\prod_i\ket{\uparrow}_i$, the Jordan-Wigner transformation
\eqref{eq:jw2} can be used to translate the eigenstates of $H_{\rs F}$ into the
eigenstates of $H_{\rs B}$. In particular, the stable four-fold ground state
degeneracy of $H_{\rs F}$ in the topological phase (guaranteed by the 10-fold
way, see Subsection~\ref{subsec:spp}) carries over to $H_{\rs B}$ \textit{as
	long as only nearest-neighbor hopping is involved} (c.f.\
Subsubsection~\ref{subsubsec:tp}). In this special case, the Fermi sea picture
describes the full spectrum of $H_{\rs B}$.
Especially, we find the ground states of $H_{\rs B}$ at the two perfectly
dimerized points: In the \textit{topological phase} with $J'=0$ and $J>0$, the
four ground states~\eqref{eq:gs2} translate to
\begin{equation}
|\Omega_{\sigma,\bar{\sigma}}\rangle 
= | \sigma\rangle_{1} | \bar{\sigma}\rangle_N\prod_{i=1}^{L-1} \left(\frac{\sigma^{-}_{2 i}
	+ \sigma^{-}_{2 i +1}}{\sqrt{2}}\right) |\uparrow\rangle_{2 i}|\uparrow\rangle_{2 i+1}
\label{eq:top}
\end{equation}
with  $\sigma, \bar{\sigma} \in \{\uparrow,\downarrow\}$.
Similarly, in the \textit{trivial phase} with $J'>0$ and $J=0$ we obtain
from~\eqref{eq:gs1}
\begin{equation}
|\Omega\rangle 
=  \prod_{i=1}^{L} \left(\frac{\sigma^{-}_{2 i-1}+ \sigma^-_{2 i }}{\sqrt{2}}\right) |\uparrow\rangle_{2 i-1}|\uparrow\rangle_{2 i}\,.
\label{eq:trivial}
\end{equation}
%

\subsubsection{Transformation of perturbations}
\label{subsubsec:tp}

In the following, we use the simplest perturbation--- next-nearest neighbor
single-particle hopping---to illustrate the consequences of the non-local
Jordan-Wigner transformation. In particular, we demonstrate the insufficiency of
the 10-fold way for our application (c.f.\ Subsection~\ref{subsec:spp}) which,
in turn, motivates the classification of interacting SPT phases described in
Subsection~\ref{subsec:spt}.

The hopping of a fermion between sites $c_1$ and $c_3$ of the same sublattice
is described as $\mathcal{O}_{\rs F}=c_1^\dag c_3+c_3^\dag c_1 $, whereas the
hopping of a hard-core boson is described by $\mathcal{O}_{\rs B}=b_1^\dag
b_3+b_3^\dag b_1$. Using the Jordan-Wigner transformation, we immediately
find the corresponding perturbations in the bosonic and fermionic model,
respectively:
\begin{subequations}
	\begin{align}
	\JW(\mathcal{O}_{\rs F} )
	&=(1-2b_2^\dag b_2)\,(b^\dag_1 b_3+b_3^\dag b_1)
	\neq\mathcal{O}_{\rs B}\,,
	\\
	\JW^{-1}(\mathcal{O}_{\rs{B}})
	&=(1-2c_2^\dag c_2)\,(c_1^\dag c_3+c_3^\dag c_1)
	\neq\mathcal{O}_{\rs F}\,.
	\end{align}
\end{subequations}
In particular, $\JW^{-1}(\mathcal{O}_{\rs B})$ and $\JW(\mathcal{O}_{\rs F})$
describe a correlated hopping of particles that can no longer be described
by a non-interacting theory.

Let us compare these two perturbations with respect to the chiral symmetry $\S$
and its bosonic counterpart $\S_{\rs B}$. It is straightforward to verify that
the perturbation $\mathcal{O}_{\rs B}$ is symmetry-allowed in the bosonic model,
i.e., $\com{\mathcal{O}_{\rs B}}{\S_{\rs B}}=0$. Since $\JW$ is an isomorphism,
we can immediately conclude that $\com{\JW^{-1}(\mathcal{O}_{\rs B})}{\S}=0$ in
the fermionic model. However, $\JW^{-1}(\mathcal{O}_{\rs B})$ is an interaction
term which is not covered by the classification of non-interacting fermions (the
10-fold way, see Subsection~\ref{subsec:spp}). In contrast, the fermionic
next-nearest neighbor hopping $\mathcal{O}_{\rs F}$ is symmetry-forbidden in the
fermionic SSH chain, and consequently we find $\com{\JW(\mathcal{O}_{\rs
		F})}{\JW(\S)}\neq 0$.

From this simple example, we can draw two important conclusions: First, the
fermionic perturbation $\mathcal{O}_{\rs F}$ violates the many-body chiral
(sublattice) symmetry $\S$ as it couples sites of the same sublattice. But it is
\textit{quadratic} and can therefore be studied in the single-particle picture
discussed in Subsection \ref{subsec:spp}. If we add this perturbation to the
Hamiltonian, $H_{\rs F}+ J'' \mathcal{O}_{\rs F}$, the latter is still described
by a single-particle matrix $\hat H$. This matrix violates the sublattice
symmetry ($U_S\hat H U_S^\dag\neq -\hat H$) and shifts the energy of the left
edge mode away from zero.

Second, the bosonic perturbation $\mathcal{O}_{\rs B}$ satisfies the symmetry
$S_{\rs B}$. But at this point we cannot conclude that this symmetry protects
our topological phase. Notably, we cannot rely on the classification of the
10-fold way since $\mathcal{O}_{\rs B}$ corresponds to a \textit{non-quadratic}
perturbation of the fermionic SSH chain. It is therefore a non-trivial but
important question whether the combination of particle number conservation and
the symmetry $S_{\rs B}$ gives rise to two distinct phases. This question is
answered by the classifications of bosonic \textit{interacting} SPT phases in
one dimension~\cite{Chen2011a,Schuch2011,Chen2011}. We proceed with this
analysis in Subsection \ref{subsec:spt}.

\subsection{Bosonic interacting SPT phases in one dimension}
\label{subsec:spt}

\subsubsection{General concept (Review)}
\label{subsubsec:gc}

Here we provide a brief review of the approach for the classification of
one-dimensional bosonic SPT phases. For an in-depth discussion, we refer the
reader to Refs.~\cite{Chen2011a,Schuch2011,Chen2011}.

Consider a one-dimensional, periodic system described by a Hamiltonian $H$ built
from local interactions that has a unique ground state $\ket{\Omega}$ and a
spectral gap $\Delta>0$ above the ground state energy $E_0$. Then it can be
shown \cite{Verstraete2006,Hastings2007,Hastings2007a,Schuch2008} that
$\ket{\Omega}$ is \textit{short-range entangled} and can be written as (or at
least approximated by) a \textit{matrix-product state (MPS)},
\begin{equation}
\ket{\Omega}
=\sum_{\{i_k\}}\,\tr{A^{i_1}A^{i_2}\dots A^{i_{L}}}\,\ket{i_1,i_2,\dots ,i_{L}}\,.
\label{eq:mps}
\end{equation}
Here, $i_k$ enumerates the local basis states in unit cell $k$. For each $i$,
$A^i$ is a matrix of dimension $D$ (called \textit{bond dimension}) so that the
coefficient of the basis state $\ket{i_1,i_2,\dots,i_{L}}$ in $\ket{\Omega}$ can
be encoded as the trace of the matrix product $A^{i_1}A^{i_2}\dots A^{i_{L}}$.
The important and non-trivial point is that for ground states $\ket{\Omega}$ of
gapped one-dimensional systems this is possible with \textit{constant} bond
dimension $D$, i.e., $D$ does not grow with the system size. Since MPS are
completely determined by their matrices $A^i$, we will write $\ket{A}$ in the
following.

Next, we focus on the Hamiltonian $H$ which, in addition, has a symmetry group
$G$ with linear representation $\rho$. Since $\com{H}{\rho(g)}=0$ for $g\in G$,
and due to the uniqueness of the ground state $\ket{\Omega}$, it follows that
the latter is invariant under the action of the symmetry up to a phase,
\begin{equation}
\rho(g)\Ket{\Omega}=\alpha(g)\ket{\Omega} \quad\text{with}\quad |\alpha(g)|=1\,.
\end{equation}
Then one can show the following \cite{Perez-Garcia2008,Chen2011}: For a symmetry
$\rho(g)=\pi(g)\otimes\pi(g)\dots\otimes\pi(g)$ that acts \textit{locally} on
each physical site via unitary representations $\pi(g)$, and an MPS $\ket{A}$
that is invariant under the action of $\rho(g)$ up to a phase,
$\rho(g)\ket{A}=\alpha(g)\ket{A}$, the matrices $A^i$ transform as
\begin{equation}
\sum_{i}[\pi(g)]_{i'i}A^i=\gamma(g)\,V^{-1}(g)\cdot A^{i'}\cdot V(g)
\label{eq:transformation}
\end{equation}
where the $\cdot$ denote matrix products and $\gamma(g)$ is a phase
(one-dimensional linear representation of $G$). Note that this necessarily
requires the invariance of the state under $\rho(g)$. If we recall that $\pi$
satisfies
\begin{equation}
\pi(g_1)\pi(g_2)=\pi(g_1g_2)
\quad\text{for}\quad g_1,g_2\in G
\label{eq:lin}
\end{equation}
as a \textit{linear} representation of $G$, one can show that the matrices $V$
in \eqref{eq:transformation} satisfy \textit{almost} the same relation
\begin{equation}
V(g_1)V(g_2)=\chi(g_1,g_2)V(g_1g_2)
\quad\text{for}\quad g_1,g_2\in G
\label{eq:proj}
\end{equation}
with $|\chi(g_1,g_2)|=1$. The function $\chi(g_1,g_2)$ is called
\textit{(2-)cocycle} or \textit{factor system} of the (projective)
representation $V(g)$; it is not arbitrary: Application of associativity,
$(g_1g_2)g_3=g_1(g_2g_3)$, yields the \textit{cocycle condition}
\begin{equation}
\chi(g_1,g_2)\chi(g_1g_2,g_3)=\chi(g_2,g_3)\chi(g_1,g_2g_3)
\label{eq:cc}
\end{equation}
which has to be satisfied in order to make \eqref{eq:proj} well-defined on the
entire group $G$.

The reason for the ``relaxed'' multiplication law \eqref{eq:proj} is that in
\eqref{eq:transformation} all phases $\chi(g_1,g_2)$ that violate the
multiplication rules of the abstract group $G$ drop out because $V(g)$ and
$V^{-1}(g)$ always pair up. Matrices that satisfy the relation \eqref{eq:proj}
realize a so called \textit{projective} representation of the group $G$ on the
bond vector space $\mathbb{C}^{D}$; \textit{linear} representations are then
special cases of projective representations with $\chi(g_1,g_2)\equiv 1$.

To proceed, recall the transformation law of MPS
matrices~\eqref{eq:transformation}:
\begin{subequations}
	\begin{align}
	\sum_{i}[\pi(g)]_{i'i}A^i
	&=\gamma(g)\,V^{-1}(g)\cdot A^{i'}\cdot V(g)\\
	&=\gamma(g)\,[f(g)V(g)]^{-1}\cdot A^{i'}\cdot [f(g)V(g)]\nonumber\\
	&=\gamma(g)\,\tilde V(g)^{-1}\cdot A^{i'}\cdot \tilde V(g)\,.
	\end{align}
\end{subequations}
Here, we defined $\tilde V(g)\equiv f(g)V(g)$ with an \textit{arbitrary}
$g$-dependent phase $f(g)$, i.e., $|f(g)|=1$ and $f(g)\in\mathbb{C}$. We can
conclude that projective representations $V$ and $\tilde V$ that are related by
a $g$-dependent phase $f(g)$ are \textit{completely equivalent} on the level of
an MPS. Now let $\chi(g_1,g_2)$ be the cocycle of the representation $V$; then
we find for the equivalent representation $\tilde V$
\begin{equation}
\tilde V(g_1)\tilde V(g_2)
=\tilde\chi(g_1,g_2)\,\tilde V(g_1g_2)
\end{equation}
with the new cocycle
\begin{equation}
\tilde\chi(g_1,g_2)=\frac{f(g_1)f(g_2)}{f(g_1g_2)}\chi(g_1,g_2)\,.
\label{eq:equivrel}
\end{equation}
This defines an equivalence relation on the set of all cocycles [i.e.,
phase-valued functions of two group elements that satisfy the cocycle condition
\eqref{eq:cc}]: Two cocycles belong to the same equivalence class, write
$\chi\sim\tilde\chi$ or $[\chi]=[\tilde\chi]$, if and only if there exists a
function $f$ such that \eqref{eq:equivrel} holds. The set of all these
equivalence classes features an abelian group structure and is called
\textit{second cohomology group of $G$ in $\group{U}(1)$}, write
$H^2(G,\group{U}(1))$. Here, the $\group{U}(1)$ encodes the fact that the
functions $f$ are phases: $f(g)=e^{i\omega(g)}$. The arguments above show that
the action of a linear representation $\rho(g)$ on a symmetric MPS is
characterized not by a \textit{particular} projective representation $V(g)$ on
its (virtual) bond space but by the \textit{cohomology class} $[\chi]\in
H^2(G,\group{U}(1))$ that its cocycle belongs to.

It is important to realize that this concept of equivalence allows for the
comparison of projective representations $V$ and $\tilde V$ even when they do
not have the same dimension $D$ because the equivalence relation
\eqref{eq:equivrel} only relies on their cocycles $\chi$ and $\tilde\chi$. Then,
the equivalence $\chi\sim\tilde\chi$ clearly does not imply that
$V(g)=f(g)\tilde V(g)$. If we keep this in  mind, we can state a crucial (and
non-trivial) fact about matrix product states \cite{Chen2011a,Schuch2011}: Let
$H_A$ and $H_B$ be two one-dimensional, gapped Hamiltonians on a common Hilbert
space $\mathcal{H}$ with symmetry $\rho(g)$ for $g\in G$ and unique ground
states $\ket{A}$ and $\ket{B}$, respectively. The latter are invariant under the
action of $\rho$ and can be written as MPS with matrices $A^i$ and $B^i$ of bond
dimensions $D_A$ and $D_B$. The action of the linear representation $\rho(g)$ on
these states induces projective representations  $V_A(g)$ and $V_B(g)$ on their
bond spaces with cocycles $\chi_A$ and $\chi_B$.

\textit{Then there exists a path $H(\lambda)$ of gapped, $\rho$-symmetric
	Hamiltonians on $\mathcal{H}$ with $H(0)=H_A$ and $H(1)=H_B$ if and only if
	$\chi_A\sim\chi_B$, i.e., iff $V_A$ and $V_B$ are projective representations
	that belong to the same cohomology class $[\chi_A]=[\chi_B]\in
	H^2(G,\group{U}(1))$.}

This implies that two symmetric states $\ket{A}$ and $\ket{B}$ belong to the
same quantum phase if and only if their corresponding cocycles (defined via
their MPS representation) are representatives of the same cohomology class. This
fact leads to the statement that the one-dimensional symmetry-protected
topological phases of interacting spin systems with symmetry group $G$ are in
one-to-one correspondence with elements of the second cohomology group
$H^2(G,\group{U}(1))$~\cite{Chen2011a,Schuch2011,Chen2011,Chen2012,Chen2013}.

\subsubsection{Application to the bosonic SSH chain with symmetry group $\group{U}(1)\times\mathbb{Z}_2^T$}
\label{subsubsec:absshcsg}

Here we apply the abstract concepts of Subsubsection~\ref{subsubsec:gc} to our
bosonic SSH chain with the relevant symmetry group
$\group{U}(1)\times\mathbb{Z}_2^T$. For simplicity, we work in the spin
language. Each unit cell $k$ contains two sites with spin $\vec\sigma_{2 k-1}$
and $\vec\sigma_{2k}$ so that there are $4$ basis states per unit cell. Then it
is straightforward to represent the ground states \eqref{eq:top} and
\eqref{eq:trivial} at the two special points ($J'=0$ or $J=0$) in terms of MPS
(here for periodic boundary conditions):

For the trivial state $\ket{B}$, we obtain the $1\times 1$ matrices
$B^{ij}_{\alpha\beta}=\sigma^x_{ij}$. Note that we introduced a double index $i
j$ with $i,j \in \{ \uparrow, \downarrow \}$ to describe the four basis states
within a unit cell. The bond dimension in the trivial state is $D=1$.

If we denote the topological state as $\ket{A}$, we obtain the matrices
\begin{equation}
A^{ij}_{\alpha\beta}\equiv\left(A^{ij}\right)_{\alpha\beta}=\delta_{i\alpha}\sigma^x_{j\beta}\,.
\end{equation}
Here, the indices $\alpha,\beta\in\{0,1\}$ label the components of the matrix
$A^{ij}$ and the bond dimension is $D=2$.

The representation of the symmetry group $\group{U}(1)\times\mathbb{Z}_2^T$ is
given by particle number conservation and the antiunitary symmetry,
\begin{equation}
R(\phi)
=\exp\left[i\phi\sum\nolimits_i\frac{\mathds{1}-\sigma_i^z}{2}\right], 
\quad
\S_{\rs B}=\prod\nolimits_i\sigma^x_i\circ K\,.
\end{equation}
\textit{Note:} In the following we label with $\S_{\rs B}$ the
\textit{representation} of the symmetry but write $S\in\mathbb{Z}_2^T$ for the
abstract group element, i.e., $\rho(S)=\S_{\rs B}$. Similarly, we write
$R(\phi)$ for the (global) \textit{representation} of $\group{U}(1)$ and
$R_\phi\in\group{U}(1)$ for the abstract group element.

On each unit cell $k$, the representations are
\begin{equation}
r_k(\phi)
=e^{i\phi}\,e^{-i\frac{\phi}{2}(\sigma_{2k-1}^z+\sigma_{2k}^z)}\,,
\quad
s_k
=\sigma^x_{2k-1}\sigma^x_{2k}\circ K\,.
\label{eq:u1loc}
\end{equation}
This is a projective representation as elements of $\group{U}(1)$ and
$\mathbb{Z}_2^T$ commute (this is implied by the direct product ``$\times$'' of
groups) whereas $s_kr_k(\phi)=e^{-2i\phi}r_k(\phi)s_k$. However, this
representation is equivalent to the linear (antiunitary) representation of the
group $\group{U}(1)\times\mathbb{Z}_2^T$
\begin{subequations}
	\begin{align}
	\tilde r_k(\phi)
	&=f(R_\phi)\,r_k(\phi)
	=e^{-i\frac{\phi}{2}(\sigma_{2k-1}^z+\sigma_{2k}^z)}\,,
	\\
	\tilde s_k&=f(S)\,s_k
	=\sigma^x_{2k-1}\sigma^x_{2k}\circ K\,,
	\end{align}
\end{subequations}
with $f(R_{\phi})=e^{-i\phi}$ and $f(S)=1$.

Now we can calculate the action of $\tilde r_k(\phi)$ and $\tilde s_k$ on the
matrices $A^{ij}$ and $B^{ij}$ of the MPS $\ket{A}$ and $\ket{B}$. Note that
both $\ket{A}$ and $\ket{B}$ are invariant under the action of $\tilde
R(\phi)=\prod_k\tilde r_k(\phi)$ and $\tilde\S_{\rs B}=\S_{\rs B}$ so that we
can derive their projective representations on the virtual bond indices. The
question is whether the corresponding cocycles belong to different cohomology
classes of $H^2(\group{U}(1)\times\mathbb{Z}_2^T,\group{U}(1))$.

For the matrices of the topological state $\ket{A}$, the action of $\tilde
r_k(\phi)$ yields
\begin{subequations}
	\begin{align}
	\tilde A^{i'j'}_{\alpha\beta}
	&=\sum_{ij}\,
	\left[e^{-\frac{\phi}{2}\sigma^z}\right]_{i',i}
	\left[e^{-\frac{\phi}{2}\sigma^z}\right]_{j',j}
	\,A^{ij}_{\alpha\beta}
	\\
	&=\left[e^{\frac{\phi}{2}(\mathds{1}-\hat\sigma^z)}A^{i'j'}
	e^{-\frac{\phi}{2}(\mathds{1}-\hat\sigma^z)}\right]_{\alpha\beta}\,.
	\end{align}
\end{subequations}
Hence we find the projective representation
$V_A(R_\phi)=e^{-\frac{\phi}{2}(\mathds{1}-\hat\sigma^z)}$ for $R_\phi\in
\group{U}(1)$.

An analogous calculation yields the action of $\tilde s_k$,
\begin{subequations}
	\begin{align}
	\tilde A^{i'j'}_{\alpha\beta}
	&=\sum_{ij}\,
	\sigma^x_{i',i}
	\sigma^x_{j',j}
	\,\left(A^{ij}_{\alpha\beta}\right)^*
	\\
	&=\left[\hat\sigma^x A^{i'j'}\hat\sigma^x\right]_{\alpha\beta}\,.
	\end{align}
\end{subequations}
Thus we have $V_A(S)=\hat\sigma^xK$ (the complex conjugation $K$ must be added
for the correct evaluation of the projective representation \cite{Chen2011}). It
is now straightforward to evaluate the corresponding cocycles
$\chi_A(R_\phi,R_\theta)=1$ and $\chi_A(S,S)=1$. However, $\chi_A(S,R_\phi)$ and
$\chi_A(R_\phi,S)$ are non-trivial. We are allowed to set $\chi_A(R_\phi,S)=1$
and obtain
\begin{subequations}
	\begin{align}
	V_A(R_\phi)\,V_A(S) 
	&=e^{-i\frac{\phi}{2}(\mathds{1}-\hat\sigma^z)}\hat\sigma^xK
	=V_A(R_\phi S)
	\\
	V_A(S)\,V_A(R_\phi)
	&=\hat\sigma^xK\,e^{-i\frac{\phi}{2}(\mathds{1}-\hat\sigma^z)}\nonumber\\
	&=\chi_A(S,R_\phi)\cdot V_A(S R_\phi)
	\end{align}
\end{subequations}
with $\chi_A(S,R_\phi)=e^{i\phi}$ as the only non-trivial value of the cocycle.

If we follow the same procedure for the trivial state $\ket{B}$ with matrices
$B^{ij}_{\alpha\beta}=\sigma^x_{ij}$, it is clear that we end up with the
trivial representation,
\begin{equation}
V_B(R_\phi)= V_B(S)=
V_B(SR_\phi)=V_B(R_\phi S)=1\,,
\end{equation}
with the trivial cocycle $\chi_B(g_1,g_2)=1$ for all
$g_1,g_2\in\group{U}(1)\times\mathbb{Z}_2^T$. Clearly $\ket{B}$ deserves
the label ``trivial'': it is a product state and its cohomology class is
$[\chi_B]=[1]\in H^2(\group{U}(1)\times\mathbb{Z}_2^T,\group{U}(1))$.

The crucial question is now whether $[\chi_A]\neq[\chi_B]$. Since $\chi_B$ is
trivial, the question is whether $\chi_A$ can be trivialized with a phase factor
$f(g)$, i.e., whether
\begin{equation}
\chi_A(g_1,g_2)
\stackrel{?}{=}
\frac{f(g_1)f^{\sigma(g_1)}(g_2)}{f(g_1g_2)}
\end{equation}
for all $g_1,g_2\in\group{U}(1)\times\mathbb{Z}_2^T$. Here, $\sigma(g)=+1(-1)$
if $g$ is represented by a unitary (antiunitary) operator (this is a consequence
of antiunitary representations and captured by \textit{twisted} group
cohomology~\cite{Yang2017,Kapustin2017}); in particular, $\sigma(R_\phi)=+1$ and
$\sigma(S)=-1$.

One can show that this is impossible: First, note that
$\chi_A(R_{\phi},R_{\theta})=1=f(R_{\phi})f(R_{\theta})/f(R_{\phi+\theta})$
implies that $f(R_\phi)$ is a linear, one-dimensional representation of
$\group{U}(1)$, i.e., it takes the form $f(R_\phi)=e^{ik\phi}$ for
$k\in\mathbb{Z}$. If we combine $\chi_A(R_\phi,S)=1=f(R_\phi)f(S)/f(R_\phi S)$
and $\chi_A(S,R_\phi)=e^{i\phi}=f(S)f^{-1}(R_\phi)/f(SR_\phi)$ and use that
$R_\phi S=S R_\phi$, we find $e^{i\phi}=1/f^2(R_\phi)=e^{-2ik\phi}$; but this
cannot be satisfied for all $\phi\in[0,2\pi)$ with $k\in\mathbb{Z}$.

We conclude that $[1]\neq [\chi_A]\in
H^2(\group{U}(1)\times\mathbb{Z}_2^T,\group{U}(1))$ and therefore $\ket{A}$
indeed is a topological phase, symmetry-protected by
$\group{U}(1)\times\mathbb{Z}_2^T$; in particular, $\ket{A}$ and $\ket{B}$
cannot be connected by a smoothly varying Hamiltonian $H(\lambda)$ that commutes
with the symmetry operators and remains gapped in the thermodynamic limit.

As a final remark, we stress that there are in total \textit{four} distinct
cohomology classes of $\group{U}(1)\times\mathbb{Z}_2^T$~\cite{Chen2013} and
\begin{equation}
H^2(\group{U}(1)\times\mathbb{Z}_2^T,\group{U}(1))
=\mathbb{Z}_2\times\mathbb{Z}_2\,.
\end{equation}
That is, the symmetry group $\group{U}(1)\times\mathbb{Z}_2^T$ can protect two
additional topological phases that smoothly connect neither to $\ket{A}$ nor to
$\ket{B}$ (these are related to antiunitary representations that square to $-1$,
i.e., $V(S)^2=-\mathds{1}$).

\subsection{Perturbative argument}
\label{subsec:pp}

Here we provide an intuitive argument based on perturbation theory that
contrasts the fragility of the non-interacting fermionic phase of $H_{\rs F}$
with the robustness of the interacting bosonic phase of $H_{\rs B}$ in the
presence of next-nearest neighbor hopping (see also $\O_{\rs F}$ and $\O_{\rs
	B}$ in Subsubsection~\ref{subsubsec:tp}).

We start from a perfectly dimerized system with $J'=0$ and $J>0$ and consider
only the leftmost edge site and its adjacent dimer. The Hamiltonian reads
$H_x=H_x^0+H_x^{\rs pert}$ with $H_x^0=-J(x_2^\dag x_3+x_3^\dag x_2)$ and
perturbation
\begin{equation}
H_x^{\rs pert}=-J'(x_1^\dag x_2+x_2^\dag x_1)-J''(x_1^\dag x_3+x_3^\dag x_1)\,.
\end{equation}
Here, $x\in\{ c,b\}$ where $c$ denotes fermions and $b$ hard-core bosons. Note
that the term proportional to $J''$ \textit{violates} the chiral symmetry $\S$
for $x=c$ but \textit{respects} the bosonic symmetry $\S_{\rs B}$ for $x=b$.  In
both cases, the degenerate ground states of $H_x^0$ have energy $-J$ and read
\begin{equation}
\ket{\Omega_m}=\left(x_1^\dag\right)^m\frac{x_2^\dag+x_3^\dag}{\sqrt{2}}\ket{0}
\end{equation}
with $m\in\{0,1\}$. If we consider $|J'|,|J''|\ll J$ as small, up to
second-order degenerate perturbation theory yields the following energy shift
for $m=0$
\begin{equation}
E^{(2)}_0 
=\frac{\left|\bra{0}x_1\,H_x^{\rs pert}\,\ket{\Omega_0}\right|^2}{-J-0}=-\frac{1}{2J}\,(-J'-J'')^2\,.
\end{equation}
This result is independent of the statistics. In contrast, for the ground state
with occupied edge mode ($m=1$), we find
\begin{equation}
E^{(2)}_1
=\frac{\left|\bra{0}x_3x_2\,H_x^{\rs pert}\,\ket{\Omega_1}\right|^2}{-J-0}
=-\frac{1}{2J}\,(-J'\pm J'')^2
\end{equation}
where the $+$ ($-$) holds for $x=c$ ($x=b$). The different signs are a
consequence of the different statistics of the operators: for bosonic operators
we find that both states are shifted by the same energy, in agreement with the
robust ground state degeneracy. In contrast, for fermionic operators the two
states acquire different energies, which leads to a lifting of the ground state
degeneracy. Note that the result for fermions can alternatively be derived from
a perturbation of the single-particle Hamiltonian, which then also implies a
splitting of the edge modes in the single-particle spectrum for bosons.

\section{Connection to the Haldane phase}
\label{sec:isotropic}

Here we demonstrate that the topological phase of the bosonic SSH chain is
smoothly connected to the antiferromagnetic spin-$1$ Heisenberg model (AFHM)
which is known to be in the famous Haldane
phase~\cite{Haldane1983a,Haldane1983,Affleck1987}. The latter is an iconic,
symmetry-protected topological phase, protected by either the dihedral group
$D_2=\mathbb{Z}_2\times\mathbb{Z}_2$ of spin rotations or time-reversal symmetry
$\mathbb{Z}_2^T$ (which explains why our system is also protected by these
symmetries).

The relation between alternating spin-$\frac{1}{2}$ chains [such as ours, see
Eq.~\eqref{eq:sshhb}] and the Haldane phase has been studied before
\cite{Yoshida1989,Hida1992a,Hida1992,Yamanaka1993,Haghshenas2014}. To relate our
model~\eqref{eq:sshhb} to an alternating spin-$\frac{1}{2}$ Heisenberg chain
(which, in turn, connects to an effective AFHM for $J'>0$ and
$J<0$~\cite{Hida1992a}), we introduce the family of Hamiltonians
\begin{equation}
\begin{aligned}
H_\delta= 
-&\frac{J'}{2}\sum_{i=1}^L (\sigma_{2i-1}^x\sigma_{2i}^x 
+ \sigma_{2i-1}^y\sigma_{2i}^y+\delta\,\sigma_{2i-1}^z\sigma_{2i}^z)
\\
-&\frac{J}{2}\sum_{i=1}^{L-1} (\sigma_{2i}^x\sigma_{2i+1}^x 
+ \sigma_{2i}^y\sigma_{2i+1}^y+\delta\,\sigma_{2i}^z\sigma_{2i+1}^z)
\end{aligned}
\label{eq:iso}
\end{equation}
with arbitrary couplings $J,J'\in\mathbb{R}$ and isotropy $\delta\geq 0$; then,
$\delta=0$ describes Hamiltonian \eqref{eq:sshhb} and $\delta=1$ its isotropic
counterpart.

\begin{figure*}[t]
	\centering
	\includegraphics[width=0.95\linewidth]{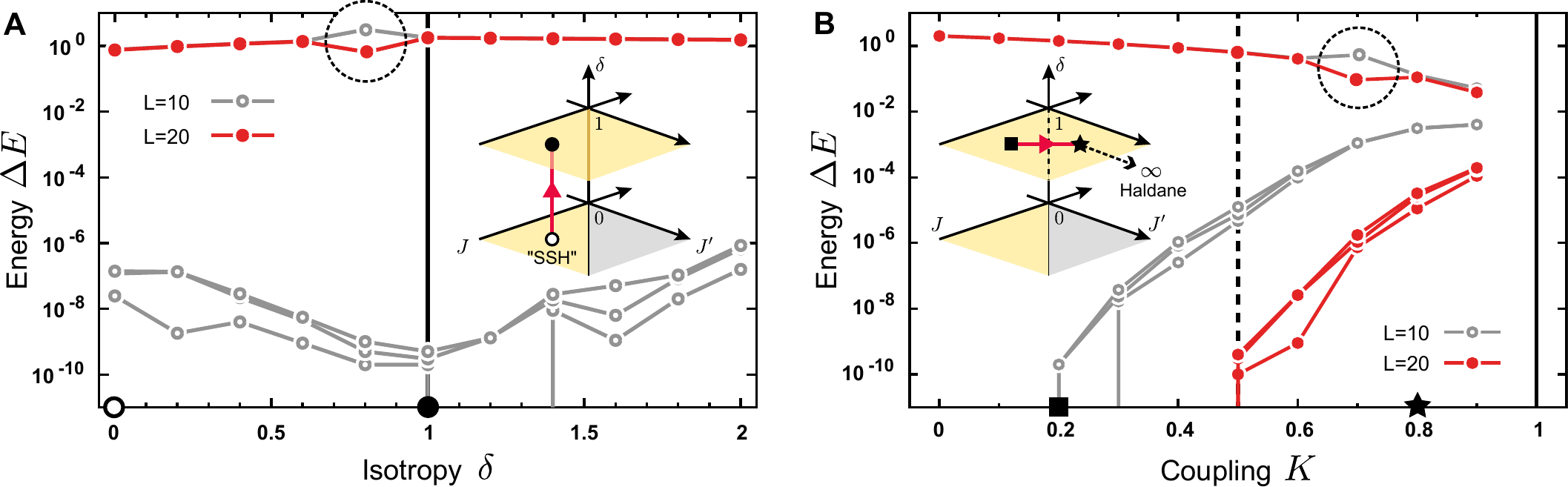}
	\caption{%
		\textbf{Path with $J'>0$ and $J<0$ from $|J'|<|J|$ and $\delta=0$ to $|J'|>|J|$ and $\delta=1$ (DMRG).}
		DMRG results for the five lowest eigenenergies of Hamiltonian \eqref{eq:iso}
		as function of the isotropy $\delta$ in (A) and the coupling parameter $K$
		in (B) for two chain lengths $L=10,20$; the ground state energy is
		normalized to zero and does not show up in the logarithmic plots. For the
		DMRG calculations (implemented with ALPS~\cite{Albuquerque2007,Bauer2011})
		we used a bond dimension of $D=300$ and $S=10$ sweeps. A four-fold ground
		state degeneracy is indicated by three energies close to zero for $L=10$
		which vanish altogether for $L=20$ as their energy is essentially zero.  The
		insets sketch the path in the $J'$-$J$-$\delta$-parameter space.
		(A) For $\delta=0$ the system is equivalent to the SSH model. For $\delta=1$
		(solid vertical line), the system becomes an isotropic Heisenberg model with
		alternating bonds. The couplings are $|J'|=0.25<1.0=|J|$ so that the phase
		is topological with a four-fold ground state degeneracy for $L\to\infty$
		(shaded yellow in the inset). The results suggest that the bulk gap stays
		open for $\delta=0\to 1$ so that the system remains in the same phase.
		(B) A path with couplings $J'=K$ and $J=K-1$ ($0\leq K\leq 1$) for the
		isotropic chain ($\delta=1$). Apparently there is no phase transition for
		$K=-J=J'=0.5$ (dashed line). This suggests that the phases for $|J'|>|J|$
		and $|J'|<|J|$ are the same if $J'>0$ and $J<0$. The limit $J'\to\infty$
		leads to an effective antiferromagnetic spin-$1$ Heisenberg chain in the
		gapped Haldane phase~\cite{Hida1992a} (inset). Note that for $K=1$ the
		ground state is a chain of decoupled triplets with extensive ground state
		degeneracy (solid line).
	}
	\label{fig:5}
\end{figure*}

\subsection{Sign-independence for $\delta=0$}
\label{subsec:isotropic}

Here we show that without interactions ($\delta=0$) all four sign-combinations
$J'\gtrless 0$ and $J\gtrless 0$ with $|J'|<|J|$ belong to the same SPT phase.
Without Ising-type interactions $\sigma_i^z\sigma_j^z$, the Hamiltonian
\eqref{eq:iso} maps onto free fermions under Jordan-Wigner transformation.
Fermions permit for gauge transformations $c_k\to \tilde c_k=e^{i\varphi_k}c_k$
with arbitrary phases $\varphi_k$; i.e., the mapping preserves the fermion
algebra $\acom{\tilde c_k}{\tilde c_l^\dag}=\delta_{kl}$. It is now
straightforward to check that this transformation translates into a unitary
mapping between equivalent representations of the Pauli algebra
\begin{equation}
\sigma_\varphi^x\equiv\cos(\varphi)\,\sigma^x-\sin(\varphi)\,\sigma^y,
\quad
\sigma_\varphi^y\equiv\sin(\varphi)\,\sigma^x+\cos(\varphi)\,\sigma^y \nonumber
\end{equation}
and $ \sigma_\varphi^z\equiv \sigma^z$ for $\varphi\in [0,2\pi)$ (in particular,
we still have
$\sigma_\varphi^a\sigma_\varphi^b=\delta^{ab}\mathds{1}+i\varepsilon^{abc}\sigma_\varphi^c$).
It follows that the commutation relations between the symmetry $\S_{\rs B}$ and
the unitarily transformed spins $\sigma_\varphi^a$ are independent of $\varphi$;
in particular, terms which commute with $\S_{\rs B}$ for $\varphi=0$ do not
violate this symmetry for $\varphi >0$ (this is also true for particle number
conservation).  We now make in \eqref{eq:iso} every other spin
$\varphi$-dependent,
\begin{equation}
\begin{aligned}
H_\delta(\varphi)= 
-&\frac{J'}{2}\sum_{i=1}^L (\sigma_{\varphi,2i-1}^x\sigma_{2i}^x 
+ \sigma_{\varphi,2i-1}^y\sigma_{2i}^y+\delta\,\sigma_{\varphi,2i-1}^z\sigma_{2i}^z)\\
-&\frac{J}{2}\sum_{i=1}^{L-1} (\sigma_{2i}^x\sigma_{\varphi,2i+1}^x 
+ \sigma_{2i}^y\sigma_{\varphi,2i+1}^y+\delta\,\sigma_{2i}^z\sigma_{\varphi,2i+1}^z)\,.
\end{aligned}
\label{eq:iso2}
\end{equation}
This defines a path of smoothly connected Hamiltonians that (i) satisfy both
symmetries and (ii) do not close the gap for $\varphi\in[0,2\pi)$ (the spectrum
of $H_\delta(\varphi)$ is $\varphi$-independent due to unitary equivalence). For
$\delta=0$ and $\varphi=0$ this yields the Hamiltonian \eqref{eq:sshhb},
$H_0(0)=H_{\rs B}$. But for $\varphi=\pi$ we find $H_0(\pi)=-H_{\rs B}$; we
conclude that the phases for $J,J'$ and $-J,-J'$ are the same. A similar
argument can be used to show that the phase diagram is mirror symmetric about
the two axes: Replacing \textit{both} spins $\sigma^a\to\sigma^a_\varphi$ on
every other \textit{dimer} (either on-site or between sites) implements the
selective sign change of either $J\to -J$ or $J'\to -J'$.

\subsection{Gapped path from the bosonic SSH chain to the spin-$1$ AFHM}
\label{subsec:gp}

We start with Hamiltonian \eqref{eq:iso} for $\delta=0$ and couplings $J'>0$ and
$J<0$ in the topological phase $|J'|<|J|$ (i.e., the ideal bosonic SSH chain
$H_{\rs B}$). For $\delta=0$, there is a four-fold ground state degeneracy and a
bulk gap that derives from the fermionic SSH chain.  In Fig.~\ref{fig:5}~(A) we
demonstrate numerically [using density matrix renormalization group methods
(DMRG)] that the gap does not close and the four-fold ground state degeneracy
remains stable if Ising-type interactions $\sigma_i^z\sigma_j^z$ are switched
on, i.e., $\delta=0\to 1$. We conclude that the bosonic SSH chain $H_{\rs
	B}=H_0$ and the alternating spin-$\frac{1}{2}$ Heisenberg chain $H_1$ belong to
the same phase for $J'>0$ and $J<0$ if $|J'|<|J|$.

Next, we calculate the spectrum of the alternating spin-$\frac{1}{2}$ Heisenberg
chain along a path with $J'=K$ and $J=K-1$ for $0\leq K\leq 1$, see
Fig.~\ref{fig:5}~(B). Again, we find a stable bulk gap and a four-fold ground
state degeneracy (up to exponential corrections in $L$). In particular, for
$|J|=|J'|$ the gap does not close (recall that $J=-J'$). This verifies that the
alternating spin-$\frac{1}{2}$ Heisenberg model undergoes \textit{no} phase
transition between $|J'|<|J|$ and $|J'|>|J|$ if $J'>0$ and $J<0$ (this is
consistent with the literature~\cite{Hida1992a,Haghshenas2014}). Note that
without interactions ($\delta=0$) there \textit{is} a phase transition from the
topological to the trivial phase (shaded gray in the insets of
Fig.~\ref{fig:5}).

From $|J'|>|J|$ one can now take the limit $J'\to\infty$. Then, the low-energy
physics is described by effective spin-$1$ degrees of freedom (triplets) formed
by pairs of spins $\vec\sigma_{2k-1}+\vec\sigma_{2k}$ in each unit cell (there
are no dangling edge spins!). These interact antiferromagnetically ($J<0$) and
give rise to an effective spin-$1$ AFHM with finite Haldane
gap~\cite{Hida1992a}.

In conclusion, we showed by numerical means that the topological phase of
$H_{\rs B}$ and the Haldane phase of the AFHM are smoothly connected via
deformations of the Hamiltonian \eqref{eq:iso2} that do not close the gap or
violate the protecting symmetries. The upshot is that the SPT phase in the main
text is essentially the Haldane phase.

\bibliography{ssh_refs_SOM}